\documentclass[11pt,a4paper]{article} 
\pdfoutput=1
\usepackage{jcappub} 
\usepackage{graphicx}
\usepackage{wasysym}

\def\beq{\begin{equation}}
\def\eeq{\end{equation}}
\def\bea{\begin{eqnarray}}
\def\eea{\end{eqnarray}}

\title{Model selection applied to reconstruction of the Primordial Power Spectrum}

\author[a,b]{J. Alberto V\'azquez}
\author[a,b]{M.~Bridges}
\author[b]{M.P.~Hobson}
\author[a,b]{A.N.~Lasenby}

\affiliation[a]{Kavli Institute for Cosmology, Madingley Road, Cambridge CB3 0HA, UK.}
\affiliation[b]{Astrophysics Group, Cavendish Laboratory, JJ Thomson Avenue, Cambridge CB3 0HE, UK.}
 
\emailAdd{jv292@cam.ac.uk}
\emailAdd{mb435@mrao.cam.ac.uk}
\emailAdd{mph@mrao.cam.ac.uk}
\emailAdd{a.n.lasenby@mrao.cam.ac.uk}

\arxivnumber{1203.1252}

\abstract{The preferred shape for the primordial spectrum of curvature
  perturbations is determined by performing a Bayesian model selection
  analysis of cosmological observations.  We first reconstruct the
  spectrum modelled as piecewise linear in $\log k$ between nodes in
  $k$-space whose amplitudes and positions are allowed to vary.  The
  number of nodes together with their positions are chosen by the
  Bayesian evidence, so that we can both determine the complexity
  supported by the data and locate any features present in the
  spectrum.  In addition to the node-based reconstruction, we consider
  a set of parameterised models for the primordial spectrum: the
  standard power-law parameterisation, the spectrum produced from the
  Lasenby \& Doran (LD) model and a simple variant parameterisation.
  By comparing the Bayesian evidence for different classes of spectra,
  we find the power-law parameterisation is significantly disfavoured by
  current cosmological observations, which show a preference for the
  LD model.}

  
\begin{document}
  \maketitle
    \flushbottom

\section{Introduction}

Cosmological inflation predicts the initial power spectrum of scalar density fluctuations to be close to
scale-invariant with just a slight scale dependence.
The simplest proposal for the shape of the spectrum is to  assume a power-law parameterisation
in terms of a spectral amplitude $A_{\rm s}$ and a spectral index or tilt parameter $n_{\rm s}$.
Although this form has been in good agreement with cosmological observations,
recent analyses from the Wilkinson Microwave Anisotropy Probe satellite (WMAP; \cite{WMAP}), and the 
Atacama Cosmology Telescope (ACT; \cite{ACT}) have confirmed that the scale invariant 
($n_{\rm s}=1$) spectrum is now excluded at $3\sigma$ C.L.
Similar results are obtained  when  measurements from the South Pole Telescope  (SPT; \cite{SPT})
 in combination  with the WMAP data  are considered.
It has also been shown that if a running of the spectral index is taken into account,
allowing deviations from the power-law spectrum,
WMAP+ ACT and WMAP+ SPT data show a preference for a negative running value at $1.8\sigma$ C.L. 
Thus, consideration of models that slightly deviate from the simple power-law might be required.
There have been several alternatives proposed.
 Some physically motivated models include
an exponential large scale cut-off \cite{Efstathiou03}, discontinuities in the early universe from 
phase transitions \cite{Barriga01}, closed universe inflation \cite{Lasenby05}, models in which the power spectrum drops 
to zero below some cut-off scale \cite{Bridle03}. Some use observational data to constrain an \textit{a priori} 
parameterisation or attempt a direct reconstruction using wavelets \cite{Mukherjee05}, 
deconvolution methods \cite{Tocchini06, Nagata10, Shafieloo10}, binning the spectrum into an arbitrary number 
of band powers \cite{Bridges06,Bridges07, Hlozek11, Guo11b}, Bayesian reconstruction \cite{Bridges08}, principle 
component  analysis \cite{Guo11}
and minimising the level of complexity needed via a 
cross-validation \cite{Peiris10}, amongst many others.
\\

In this paper, we are interested in selecting the preferred shape for the primordial spectrum 
using the Bayesian evidence as an implementation of Occam's razor: a simpler model 
should be preferred, unless the data require a more sophisticated model.
First, we determine the structure of the primordial power spectrum using an optimal model-free 
reconstruction. 
Our approach considers possible deviations from the power-law parameterisation by modelling the
spectrum as a linear interpolation between a set of `nodes' which can vary in both amplitude and $k$-position.
Within this analysis we have included phenomenological features which might be relevant to 
the description of CMB observations, such as a large scale cut-off, a broken spectrum and a 
spectrum with a  possible turn-over at any position in $k$-space.  
The reconstruction process is essentially the same {\it binning} format used previously by a 
number of authors, however here we allow the data to decide the level of complexity of the
model -- the number of nodes and their optimum position --
via the Bayesian evidence.
 Then, for comparison, we compute the Bayesian evidence for 
 a set of existing model proposals: a power-law parameterisation including both tilt and running parameter, 
 a modified power-law spectrum to account for a drop off at large scales and 
 the Lasenby \& Doran (LD) model based on a closed universe.
 Finally, for each model we compare its Bayesian evidence and 
 according to the Jeffreys guideline we select the best model preferred by current data. 
 In a previous  paper \cite{Vazquez11}, we have constrained the parameter space  which describes  
 the primordial spectrum derived  from  the  LD model. In this  work, using an  optimal model-free
 reconstruction, the  shape  of the  spectrum is determined directly from the data.
 \\

The paper is organised as follows: in the next Section we describe
the basic parameter estimation and model selection, we also list the datasets
and the cosmological parameters considered. In Section~\ref{sec:Pk} we carry out the
 reconstruction for the primordial power spectrum and also consider different 
 existing parameterisations suggested to describe the form of the spectrum.
We show the resulting parameter constraints and the evidence for each worked model. 
Finally, in Section~\ref{sec:results} we decide which model provides the
best description for current observational data and present our conclusions.

\section{Bayesian Inference}
\label{sec:Model}

\subsection{Parameter Estimation}

\noindent
Given a model or hypothesis $H$ for some data $\mathbf{D}$, Bayes' theorem tells us how to determine the probability 
distribution of the set of parameters $\mathbf{\Theta}$ on which the model depends.
  Bayes' theorem states that

\begin{equation}\label{eq:bayes}
  \Pr( \mathbf{\Theta}|\mathbf{D},H)= 
\frac{\Pr(\mathbf{D}|\mathbf{\Theta},H) \,\, \Pr(\mathbf{\Theta}|H)}{\Pr(\mathbf{D}|H)},
\end{equation}

\noindent
where, for brevity, we denote $\Pr(\mathbf{\Theta}|H) \equiv
\pi(\mathbf{\Theta})$ as the {\it prior} probability which is
upgraded through the {\it likelihood}
$\Pr(\mathbf{D}|\mathbf{\Theta},H) \equiv \mathcal{L}(\mathbf{\Theta})$
when experimental data $\mathbf{D}$ are considered.
The aim for parameter estimation is to obtain the posterior probability
$\Pr(\mathbf{\Theta}|\mathbf{D},H)$ which represents the state of
knowledge once we have taken into account the new information. The
normalisation constant is usually refered to the {\it Bayesian evidence} 
$\Pr(\mathbf{D}|H) \equiv \mathcal{Z}$.
Since this quantity is independent of the parameters $\mathbf{\Theta}$, it is often ignored in parameter
estimation but plays the central role for model selection.
\\

Throughout the analysis we compute posterior probabilities for each model 
in the light of temperature and polarisation measurements  from the 
7-year data release of  the Wilkinson Microwave Anisotropy Probe  (WMAP; \cite{WMAP}). 
  The ACT and SPT data provide similar constraints on the primary cosmological parameters
  \cite{ACT,SPT}, thus since our results are not likely to be significantly affected according
  to which is chosen, and to avoid possible problems due to overlap of the sky regions, we have
  restricted the current analysis to use of just the ACT data.
In addition to CMB data,  we include distance measurements from the
Supernova Cosmology Project Union 2  compilation (SCP; \cite{SCP}) and large scale structure data from 
the Sloan Digital Sky Survey (SDSS) Data Release 7 (DR7) Luminous Red 
Galaxy (LRG) power spectrum \cite{LRG}. We also impose a Gaussian prior from  measurements 
of the Hubble parameter today $H_0$  from the  Hubble Space Telescope (HST; \cite{HST}) key project.
\\

We consider purely Gaussian adiabatic scalar perturbations and neglect tensor contributions. 
We assume a flat $\Lambda$CDM model 
\footnote{Except for the LD model, which is based on a closed universe $\Omega_k <0$.}  specified by
the following parameters: the physical baryon $\Omega_{\rm b} h^2$ and cold dark matter density
 $\Omega_{\rm DM } h^2$ relative to the critical density ($h$ is the  dimensionless Hubble parameter 
 such that $H_0=100h$ kms$^{-1}$Mpc$^{-1}$), $\theta$ is $100 \times$ the ratio of the sound horizon 
 to angular diameter distance at last scattering surface and $\tau$ denotes the optical depth at reionisation.
Aside from the Sunyaev-Zel'dovich (SZ) amplitude $A_{SZ}$ used by WMAP analyses, 
the 148 GHz ACT likelihood incorporates two additional secondary parameters:
the total Poisson power $A_p$ at $l=3000$ and the amplitude of the clustered power $A_c$.
 The parameters describing the primordial spectra in each model are listed in
Section \ref{sec:Pk}, together with their flat priors imposed in our Bayesian analysis.

\subsection{Model Selection} 

\noindent
Model selection applies a similar type of analysis to parameter 
estimation but now at the level of models rather than parameters.
The key quantity to bear in mind is the Bayesian evidence.
 It balances the complexity of cosmological models and then, naturally, incorporates
Occam's razor. 
Let us consider several models or hypotheses $H_i$, each of them with prior probability $\Pr(H_i)$. 
Bayes' theorem for model selection is
\\

\begin{equation}\label{eq:bay_model}
 \frac{ \Pr(H_i |\mathbf{D})}{\Pr(H_j |\mathbf{D})}= 
 \frac{\Pr(\mathbf{D}| H_i)}{\Pr(\mathbf{D}| H_j)}\frac{\Pr(H_i)}{\Pr(H_j)}.
\end{equation}

\noindent
The left-hand side denotes the relative probability of the model given the
data, whereas in the right-hand side appears the key quantity to compute: the Bayesian evidence
for each model.
This evidence is nothing more than the average of the likelihood over the prior

\begin{equation}\label{eq:Evi}
  \mathcal{Z}= \int \mathcal{L}(\textbf{$\Theta$})\pi(\textbf{$\Theta$})d^N\textbf{$\Theta$},
\end{equation}

\noindent
where $N$ is the dimensionality of the parameter space.
Note that a complex model will usually lead to a higher likelihood, but if the fit is nearly 
as good as a simple model, the evidence will favour the simpler model through the smaller prior volume.
\\
 
\noindent
For  convenience, the ratio of two evidences $ \mathcal{Z}_0/ \mathcal{Z}_1$ 
(or  equivalently the  difference  in log  evidences  $ \ln \mathcal{Z}_0 -\ln \mathcal{Z}_1$) is often 
termed the Bayes factor $\mathcal{B}_{0,1}$:

\begin{equation}\label{eq:bayes}
\mathcal{B}_{0,1} =
\,\,\ln \frac{ \mathcal{Z}_0}{ \mathcal{Z}_1}.
\end{equation}
Then, the quantity $ \mathcal{B}_{0,1}$ provides an idea on how well model $0$ may fit
the data when is compared to model $1$. Jeffreys provided a suitable guideline scale on 
which we are able to make qualitative conclusions (see Table \ref{tab:Jeffrey}).
In this  work, we refer to positive (negative) values of $ \mathcal{B}_{0,1}$ as the $0$  model
being favoured (disfavoured) over  model $1$.
\\

\begin{table}
\begin{center}
\caption{Jeffreys guideline scale for evaluating the strength of evidence when two models are compared.}
\begin{tabular}{cccc} 
\cline{1-4}\noalign{\smallskip}
\vspace{0.2cm}
$| \mathcal{B}_{0,1}|$ & Odds & Probability & \,\,\, Strength\\

\hline
\vspace{0.2cm}
$<$ 1.0 &  $< $ 3 : 1  		& $<$ 0.750   & Inconclusive \\
\vspace{0.2cm}
1.0-2.5       &   $\sim$ 12 : 1      		   & 0.923           & Significant \\
\vspace{0.2cm}
2.5-5.0      &    $\sim$ 150 : 1   		    & 0.993           & Strong \\
\vspace{0.2cm}
$>$ 5.0      &       $>$ 150 : 1  	   & $>$ 0.993           & Decisive\\
\hline
\hline
\end{tabular}
\label{tab:Jeffrey}
\end{center}
\end{table}

The computation of the integral in Equation (\ref{eq:Evi}) is a very
computationally demanding process, since it requires a
multidimensional integration over the likelihood and prior.
To carry out the exploration of the cosmological parameter space
we  first modify the {\sc CAMB} code \cite{CAMB} in order to input an arbitrary  shape  of the primordial power 
spectrum\footnote{A modified {\sc CAMB} code version which allows a Node-based Parameterisation
for the primordial power spectrum 
is a available at  \url{http://www.mrao.cam.ac.uk/software/}},
 then we incorporate into the {\sc CosmoMC} software \cite{Cosmo} a substantially improved and 
fully-parallelized version of the {\it nested sampling} algorithm called {\sc MultiNest} \cite{Multi1,  Multi2}.
The {\sc MultiNest} algorithm increases the sampling efficiency for calculating the evidence and 
allows one to obtain posterior samples even from distributions with multiple modes and/or pronounced
degeneracies between parameters.

\section{Primordial Power Spectrum Parameterisation}
\label{sec:Pk}

\subsection{Power spectrum reconstruction I}
\label{sec:rec_1}

First, we perform a reconstruction for the shape of the primordial spectrum. We parameterise 
$\mathcal{P_R}(k)$ with a specific number of bins, logarithmically
spaced in $k$, and varying only each amplitude, denoted $A_{{\rm s},k_i}$. 
Throughout, we assume that most of the current relevant
information is encompassed within the scales $k_{\rm min}
=0.0001$ ${\rm Mpc}^{-1}$ and $k_{\rm max}=0.3$ ${\rm Mpc}^{-1}$, where
the combined WMAP+ACT  data significantly improves  the parameter  constraints.
Outside of these boundaries we take the spectrum to be constant with values 
equal to those at $k_{\rm min}$ and $k_{\rm max}$ respectively.
 We allow variations in the spectral amplitudes 
 with a conservative prior $A_{{\rm s},k_i}\in [1,50] \times 10^{-10} $. 
\\

\noindent
To maintain continuity between $k$-nodes, a linear  interpolation is performed such that
the form of the power spectrum is described by 
\begin{eqnarray} 
\mathcal{P_R}(k) = \left\{ \begin{array}{cc}

A_{{\rm s},k_{\rm{min}}} 				& \quad	k\le k_{\rm min}\\ 
A_{{\rm s},k_i}			      		& \quad	k \in \{k_i\}   \\ 
A_{{\rm s},k_{\rm {max}}}	 			&\quad 	k\ge k_{\rm max} \\
\end{array} \right.&& \\ \nonumber  \\
{\rm and\,\, with \,\,linear\,\, interpolation  \,\, for} \qquad
 k_{\rm min}\le k_i<&k&<k_{i+1}\le k_{\rm max}. \nonumber 
\end{eqnarray} 

\begin{figure}[t!]
\begin{center}$
\begin{array}{cc}
(a) \,\,  \mathcal{B}_{1,1} =0.00 \pm 0.30     \\
\includegraphics[trim = 1mm  2mm 5mm 0mm, clip, width=5cm, height=3.5cm]{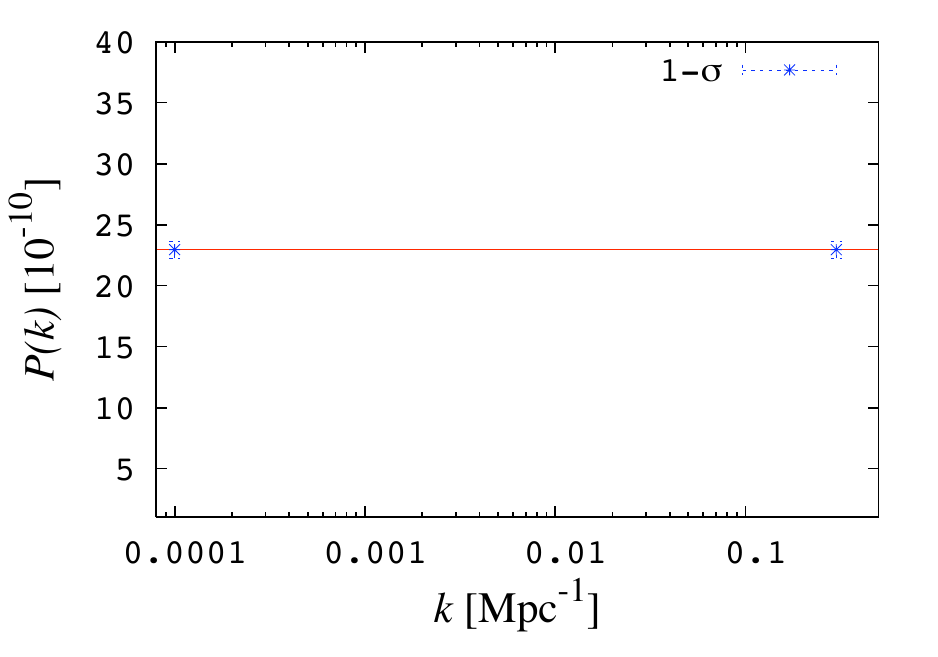}  
\includegraphics[trim = 30mm  112mm 30mm 110mm, clip, width=5cm, height=3.5cm]{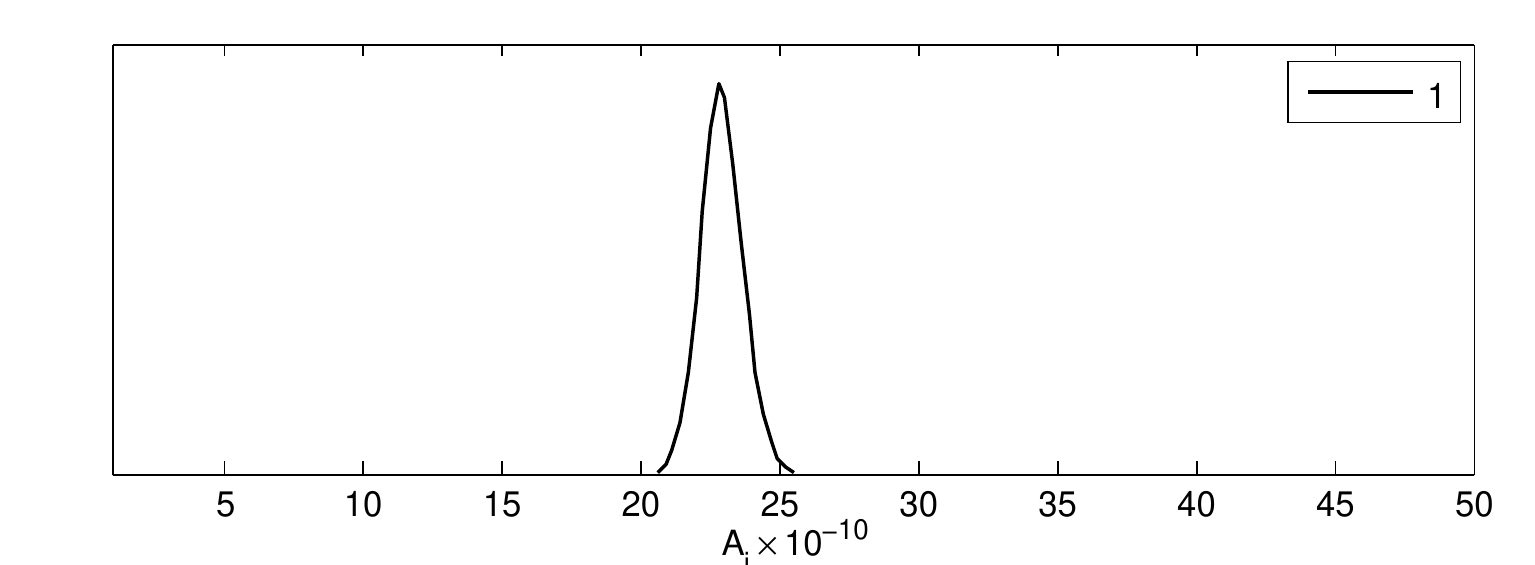}\\
(b) \,\, \mathcal{B}_{2,1} =+2.93 \pm 0.30 &\\
\includegraphics[trim = 1mm  2mm 5mm 0mm, clip, width=5cm, height=3.5cm]{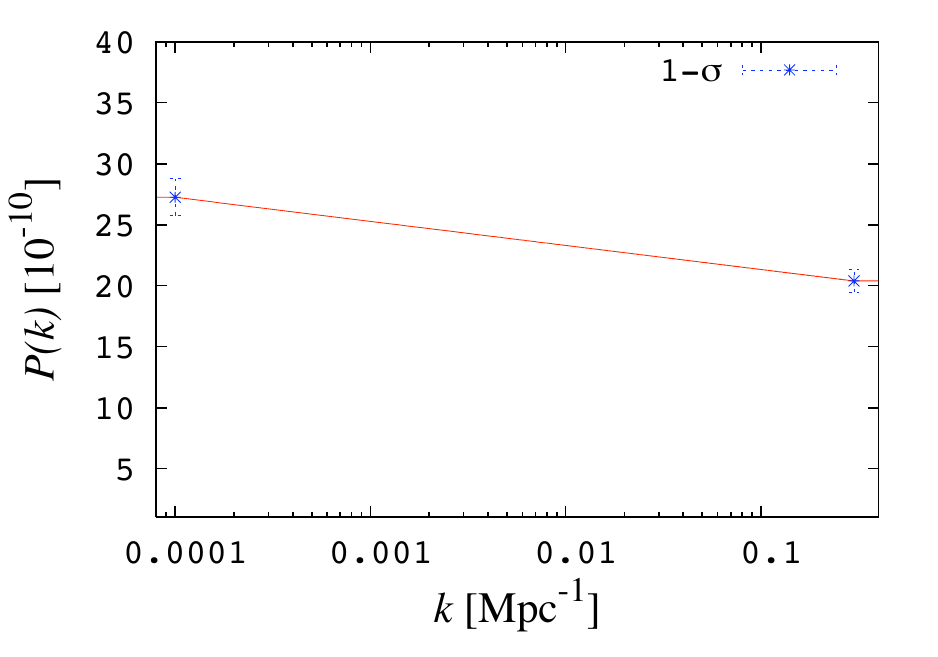}
\includegraphics[trim = 30mm  112mm 30mm 110mm, clip, width=5cm, height=3.5cm]{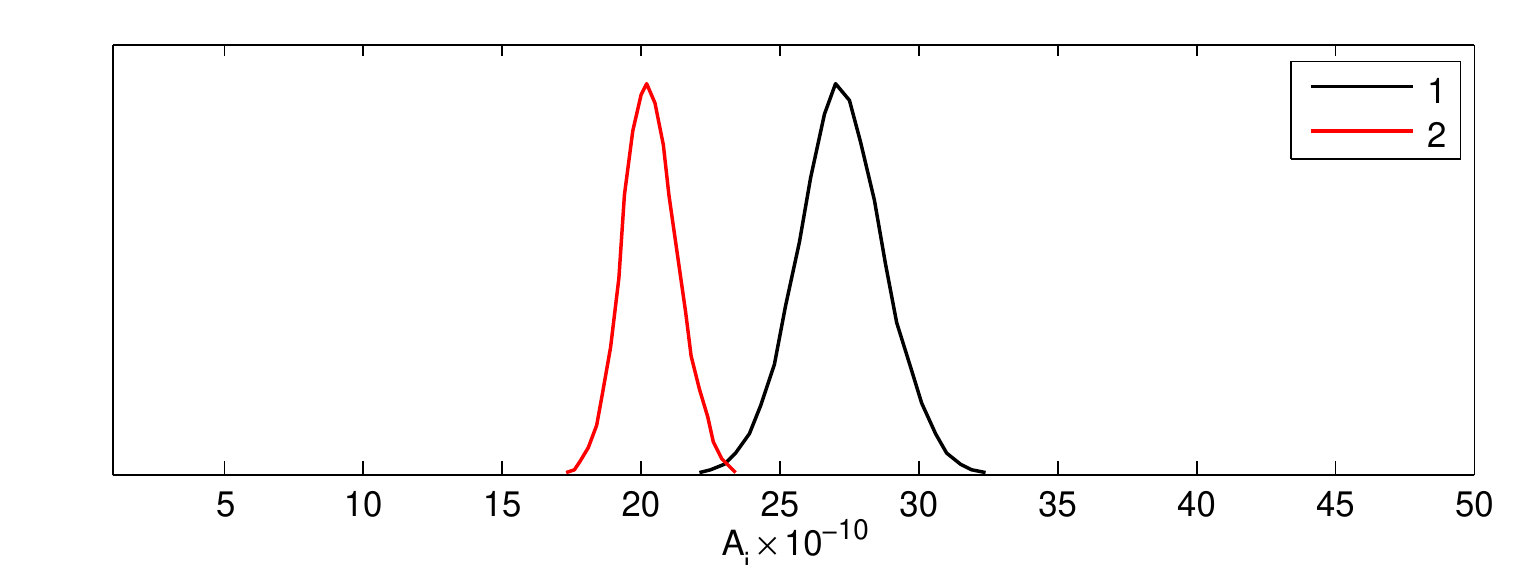}  \\
 (c) \,\, \mathcal{B}_{3,1} =+2.75 \pm 0.30  &  \\
\includegraphics[trim = 1mm  2mm 5mm 0mm, clip, width=5cm, height=3.5cm]{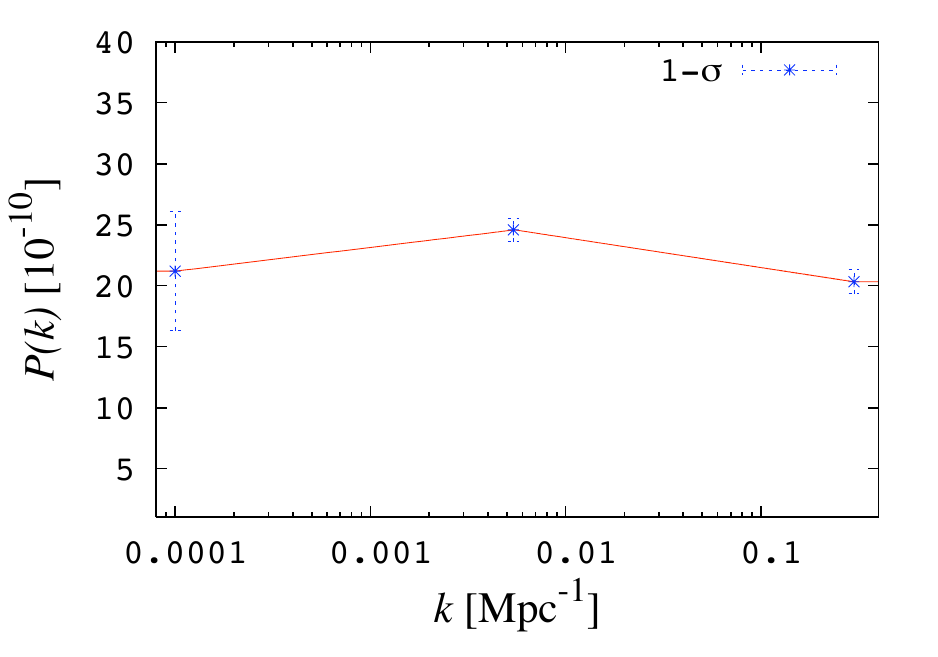} 
\includegraphics[trim = 50mm  95mm 50mm 100mm, clip, width=5cm, height=3.5cm]{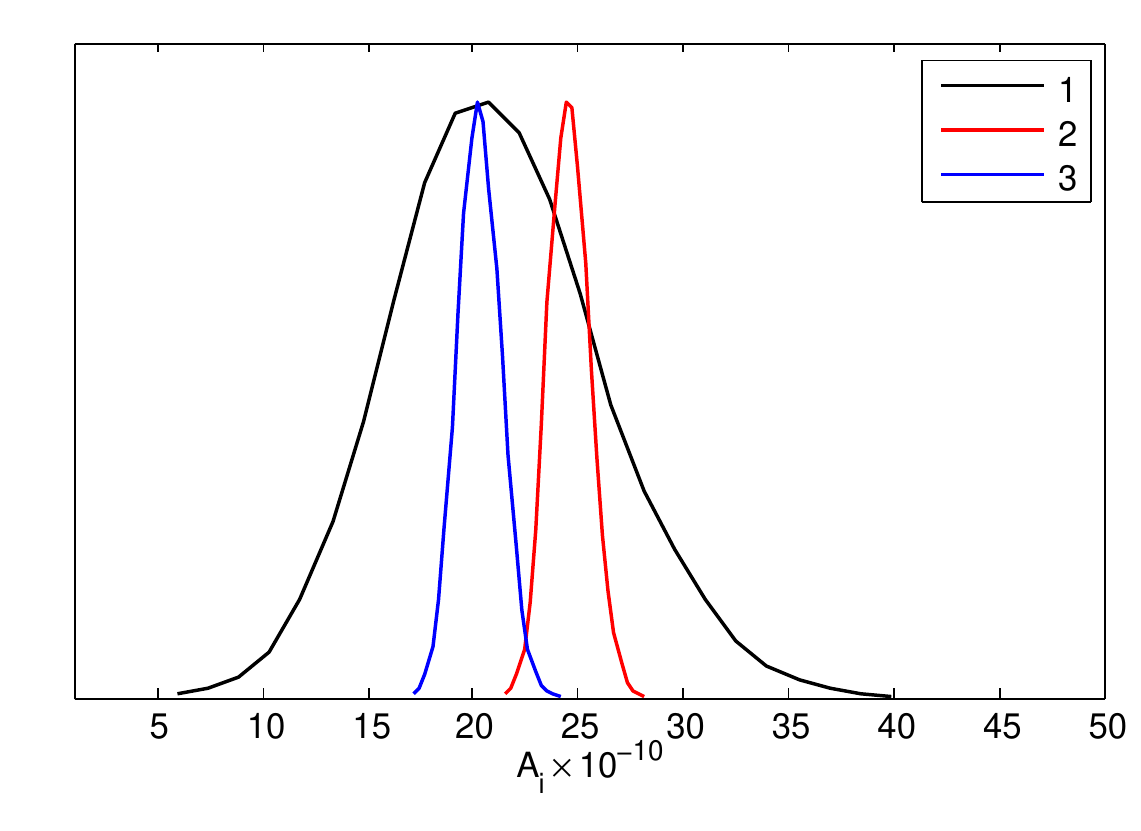}\\
 (d) \,\, \mathcal{B}_{4,1} =+0.67 \pm 0.30 &\\
\includegraphics[trim = 1mm  2mm 5mm 0mm, clip, width=5cm, height=3.5cm]{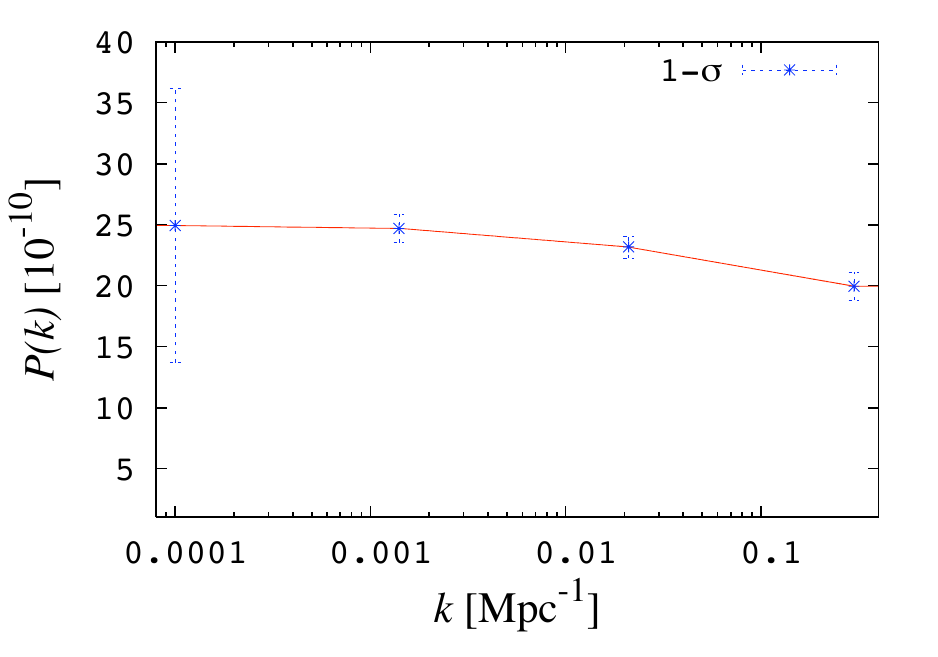} 
\includegraphics[trim = 30mm  110mm 30mm 110mm, clip, width=5cm, height=3.5cm]{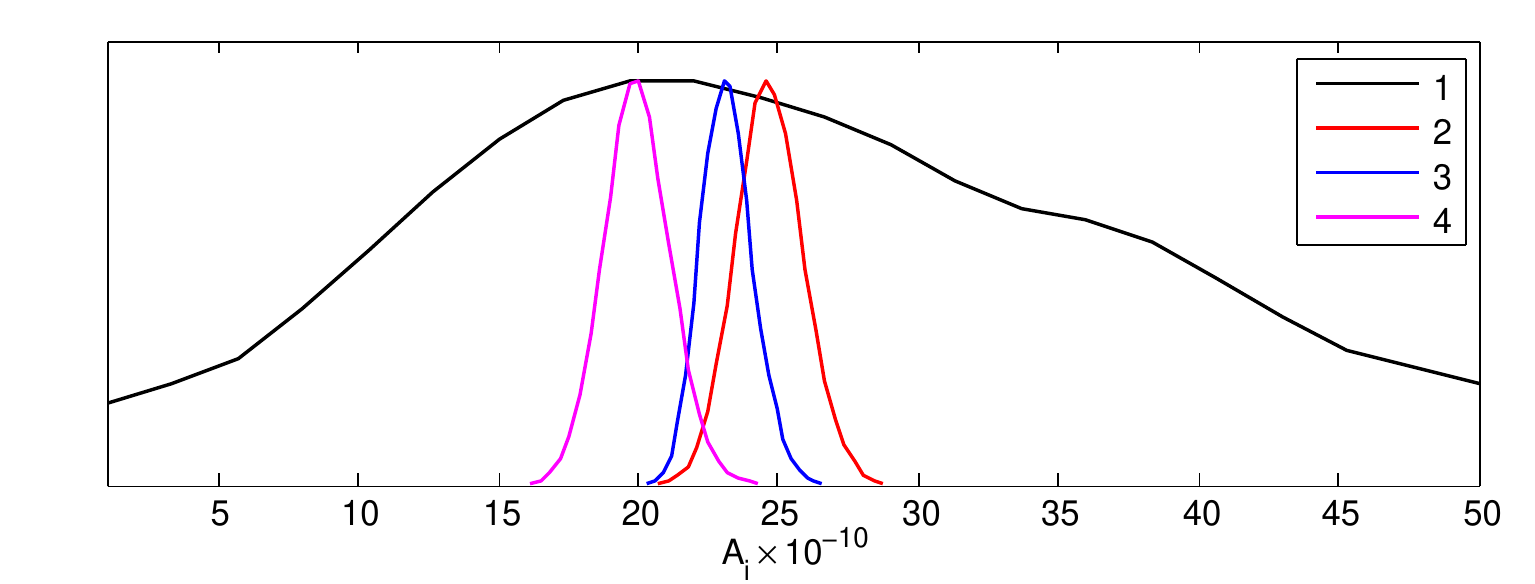} 
\end{array}$
\end{center}
\caption{Left: Reconstruction of  the primordial spectrum modelled as piecewise linear between nodes 
with fixed wavenumber $k_i$, along  with mean
amplitude  values and their corresponding $1\sigma$ error bars.
On large scales the power spectrum is constrained by WMAP data, whereas at small scales ACT/LRG
data yield tight constraints up to $k= 0. 3$ ${\rm Mpc}^{-1}$. 
Right: 1D  marginalised posterior  distribution  of the  amplitudes $A_i$
at  each bin in  each  reconstruction. The top label  in each  panel  denotes the associated Bayes factor  respect
to the base model (HZ) (a).  }
\label{fig:recons_1}
\end{figure}

We start our reconstruction by considering  
the base model is equivalent to the Harrison-Zel'dovich (HZ) spectrum ($n_{\rm s}=1$)
where the spectral amplitude $\mathcal{P_R}(k)=A_{{\rm s}}$ is the only parameter,  
see Figure \ref{fig:recons_1}.
The next model, (b), allows for two amplitudes located at $k_{\rm min}$ and $k_{\rm  max}$
to vary independently, thus emulating a tilted spectrum. 
We then add a third point (c) placed midway between the two existing nodes in (b). 
This model mimics a degree of spectral running by allowing slight variations
in the interpolated slopes between the three nodes.
Since these amplitudes are independent of each other, however, 
there is no need to pick any particular pivot point as in the case of power-law
parameterisation, hence providing more freedom in the shape of the spectrum.
The presence of   a turn-over in the global structure of the resultant spectrum,
 shown in Figure \ref{fig:recons_1} (c), points to some deviation from a simple tilt.
We might continue adding nodes in this fashion until some arbitrary accuracy of
model fit is achieved, but  always bearing in mind that the inclusion of new unnecessary nodes 
is penalised through the Bayesian evidence.
We, then consider a fourth stage where the $k$-space is logarithmically split 
into three equally spaced regions, (d). 
At small scales the shape of the power spectrum is well constrained with tight
 error bars on each node, whereas on large scales the error bars tell us  
there is still room for new features (within the limited amount of information due to cosmic  variance). 
Notice the increased error bars due to the addition of an arbitrary
number of band-powers and correlations created between them.  We also
observe the evidence has dropped for four $k$-nodes, therefore this
stage seems to be a reasonable point to stop adding parameters in the
reconstruction process.
Figure \ref{fig:recons_1} illustrates the corresponding form of the reconstructed 
spectra from the mean posterior estimates (with $1\sigma$ error bars on its corresponding amplitudes), together
with 1D marginalised posterior distribution for the amplitude at  each node  and  for each
reconstruction. In each model we include the Bayes factor compared to the
base model (HZ).
 \\
 
 The reconstructed spectra are assessed according to the Jeffreys guideline shown in Table \ref{tab:Jeffrey}.
The Bayesian evidence
between the base model and the two-node model  $\mathcal{B}_{2,1} =+2.93 \pm 0.30$ points out that the 
HZ is strongly disfavoured when compared to a tilted spectrum, in agreement with WMAP/ACT \cite{ACT} results.
The addition of complexity in the third stage provides more flexibility in the shape of the reconstructed 
spectrum. 
  The evidence between the  two-node and three-node model, $\mathcal{B}_{2,3} =+0.18 \pm 0.30$,
  is  too small to  draw  any  decisive  conclusion, though the  evidence marginally
  prefers the simple tilted spectrum.
Although the  reconstructed shape of the spectrum in the fourth stage is similar to the one 
obtained in the  second stage,  the  four $k$-node model is penalised 
 because of the inclusion of unnecessary information.
   Thus, the peak of the evidence at model (b) shows that the optimal reconstruction 
   contains, surprisingly, just  two nodes, as is shown in Figure \ref{fig:recons_1}.
 Therefore, according to this  reconstruction process,  parameterisations such as the HZ and those containing 
 more than three  $k$-nodes are hence disfavoured by current observations.
   At  this point  of the  analysis, with fixed $k$-node positions, our results are consistent with those  
   obtained  by  \cite{Guo11b}, where according to  the Akaike  information criterion,  the  preferred model
    is given by a two-node  spectrum.    
    \\
    
    To extract the global structure of the spectrum we have carried out a  reconstruction process by
    placing nodes at particular positions in the $k$-space.
    However, to localise features in $k$-space, we may
   consider moving either back or forth the internal $k$-nodes until
   we find their optimal position; we reconsider this option in an
   improved method in the next Section.

\subsection{Power spectrum reconstruction II}

In Section \ref{sec:rec_1} we reconstructed the primordial spectrum using a  standard binning
process: fix $k$-node positions and vary only the amplitudes. 
We now consider  a reconstruction of the spectrum where the  internal $k$-node positions
vary, as well as their amplitudes.

\subsubsection{Running-like spectrum}
\label{sec;k_i}

In order to look for deviations from the simple power-law model, we
consider a model with two fixed $k$-nodes at sufficiently separated positions $[k_{\rm min},k_{\rm max}]$,
with varying amplitudes 
[$A_{{\rm s},k_{\rm  min}}$, $A_{{\rm s},k_{\rm  max}}$], and place inside
additional `nodes' with the freedom to move around in both position
$k_i$ and amplitude $A_{{\rm s},k_i}$.
Despite the simplicity of this approach, it covers a large variety of shapes for the primordial
spectrum. The freedom of the position  of the internal $k$-nodes allows us to localise the 
best position for a turn-over (if any) and
the amplitudes are able to describe the global structure of the spectrum. 
\\

Analogously to Section \ref{sec:rec_1}, we have maintained the same priors for the
spectral amplitude  $A_{{\rm s},k_i}=[1,50] \times 10^{-10} $, whereas  on the  $k$-position
we restrict to the  physical prior $\log k_i =[\log k_{\rm min},\log k_{\rm max} ]$.
Hence, for this type of nodal-reconstruction the spectrum  is described  by
\begin{eqnarray} 
\mathcal{P_R}(k) = \left\{ \begin{array}{ll} 

A_{{\rm s},k_{\rm min}} 				& \quad k\le k_{\rm min}\\ 
A_{{\rm s},k_i }			      		& \quad	k_{\rm min}< k_i<k_{i+1}< k_{\rm max}   \\ 
A_{{\rm s},k_{\rm max}}	 			&\quad 	k\ge k_{\rm max} 
\end{array} \right. && \\ \nonumber \\ 
{\rm and\,\, with\,\,  linear\,\,interpolation\,\, for \quad} 
  k_{\rm min}\le &k_i&\le k_{\rm max}. \nonumber 
\end{eqnarray} 




\begin{figure}
\begin{center}$
\begin{array}{cc}
 (k_1) \,\, $$\mathcal{B}_{k_1,1} =+4.26 \pm 0.30$$ &\\
 \includegraphics[trim = 1mm  2mm 5mm -5mm, clip, width=5cm, height=4cm]{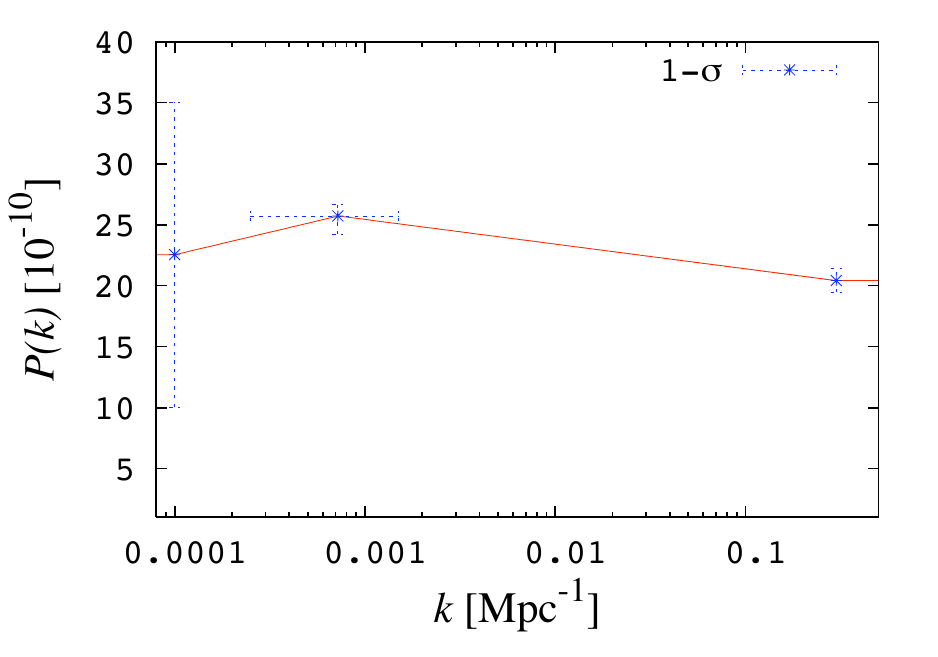}
\includegraphics[trim = 30mm  75mm 45mm 70mm, clip, width=5cm, height=4cm]{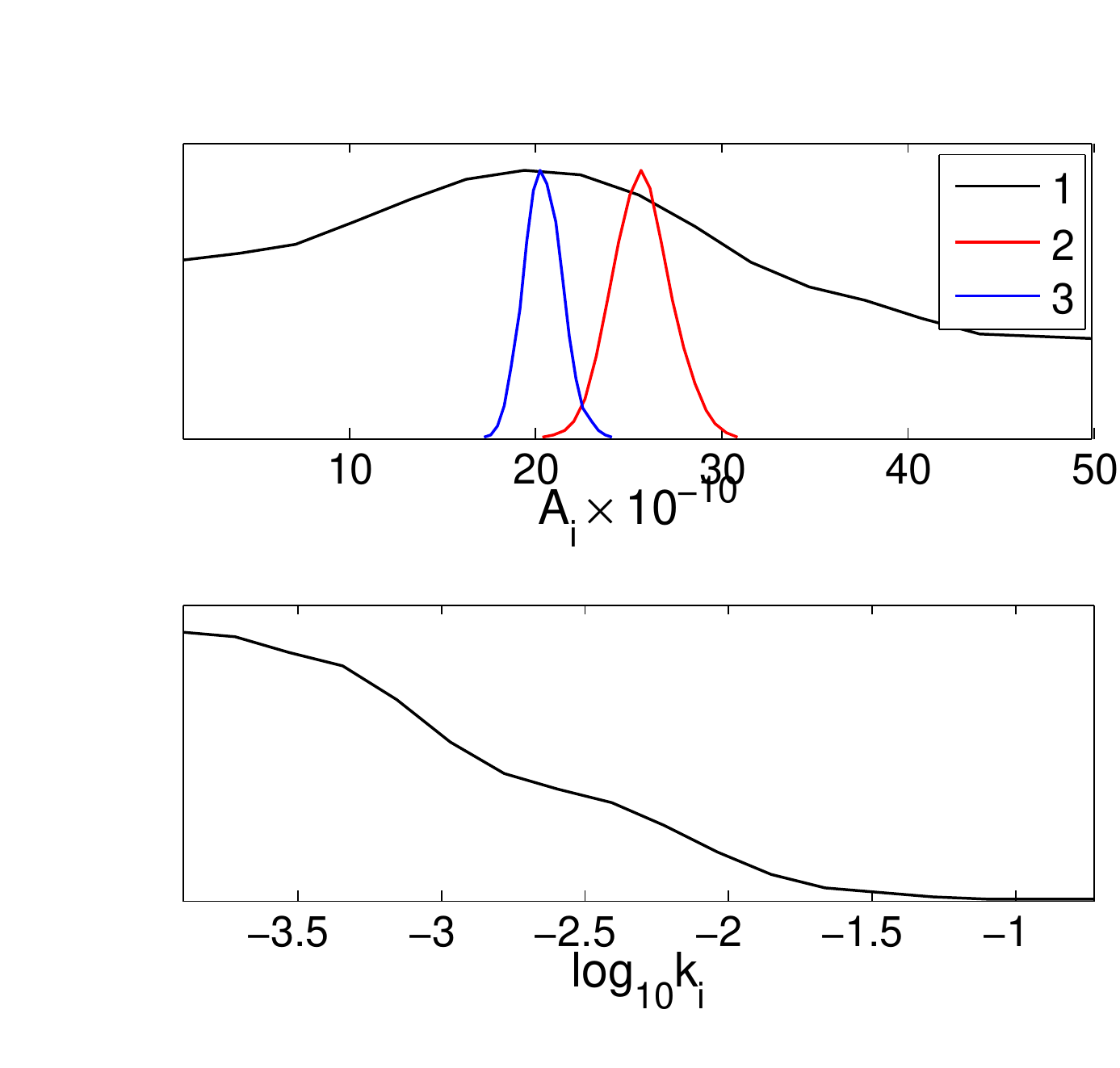} \\
 (k_2) \,\, $$\mathcal{B}_{k_2,1} =+3.73 \pm 0.30$$ &\\
 \includegraphics[trim = 1mm  2mm 5mm -5mm, clip, width=5cm, height=4cm]{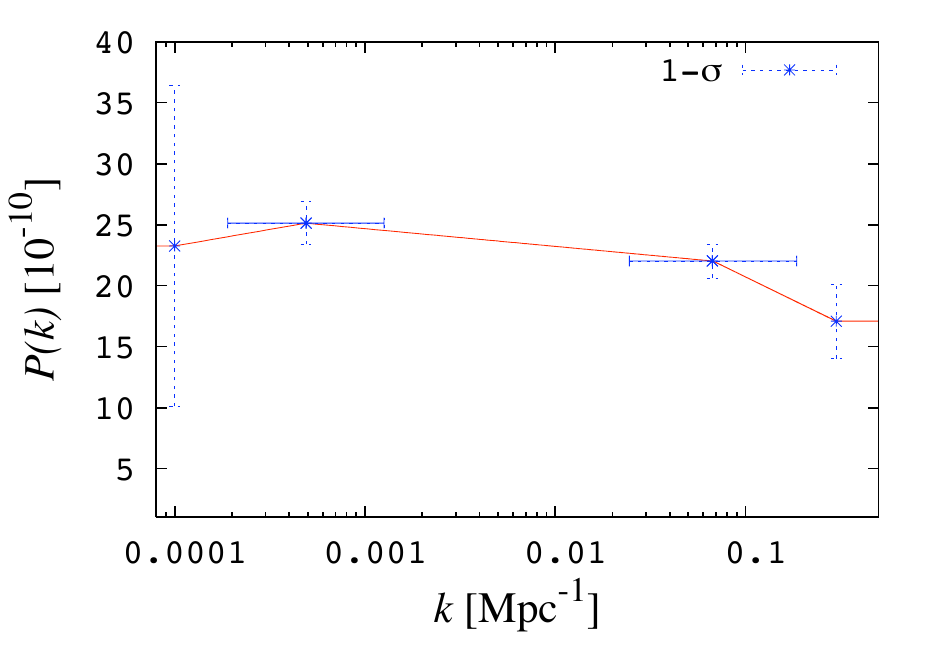}
\includegraphics[trim = 30mm  65mm 45mm 65mm, clip, width=5cm, height=4cm]{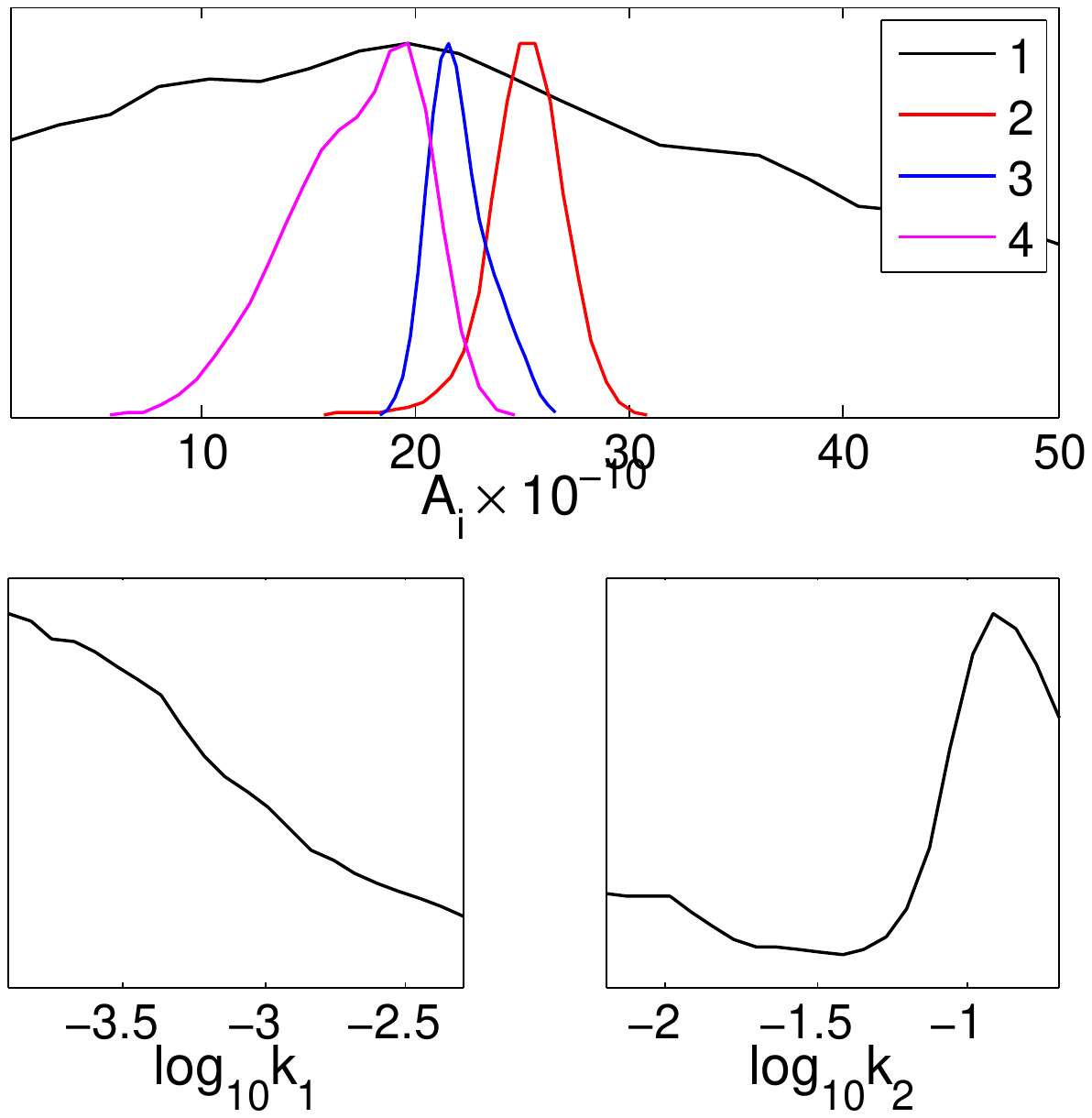} \\
 (k_3) \,\, $$\mathcal{B}_{k_3,1} =+3.49 \pm 0.30$$ &\\
 \includegraphics[trim = 1mm  2mm 5mm -5mm, clip, width=5cm, height=4cm]{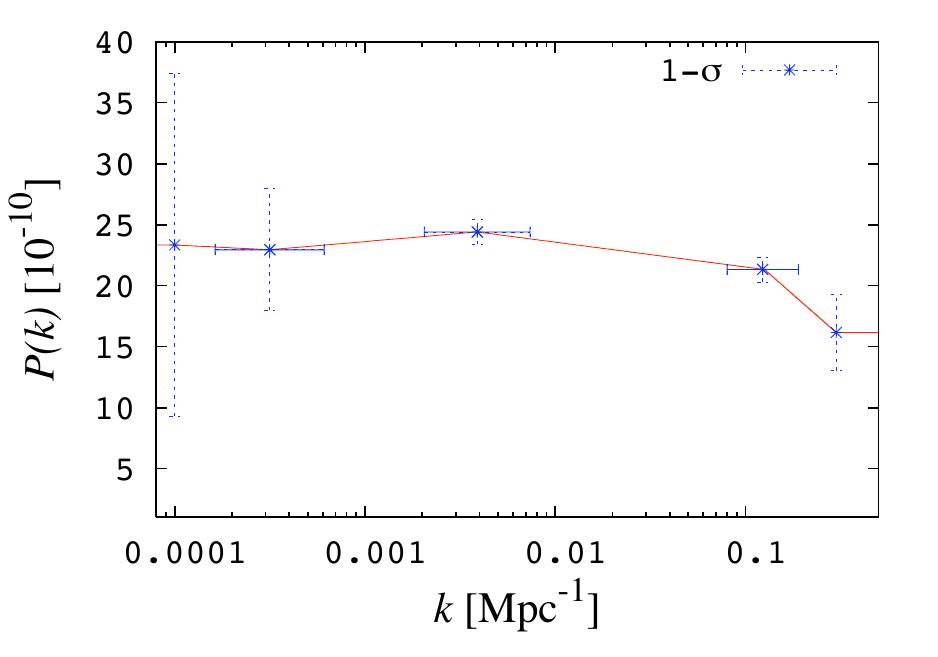}
\includegraphics[trim = 30mm  80mm 35mm 78mm, clip, width=5cm, height=4cm]{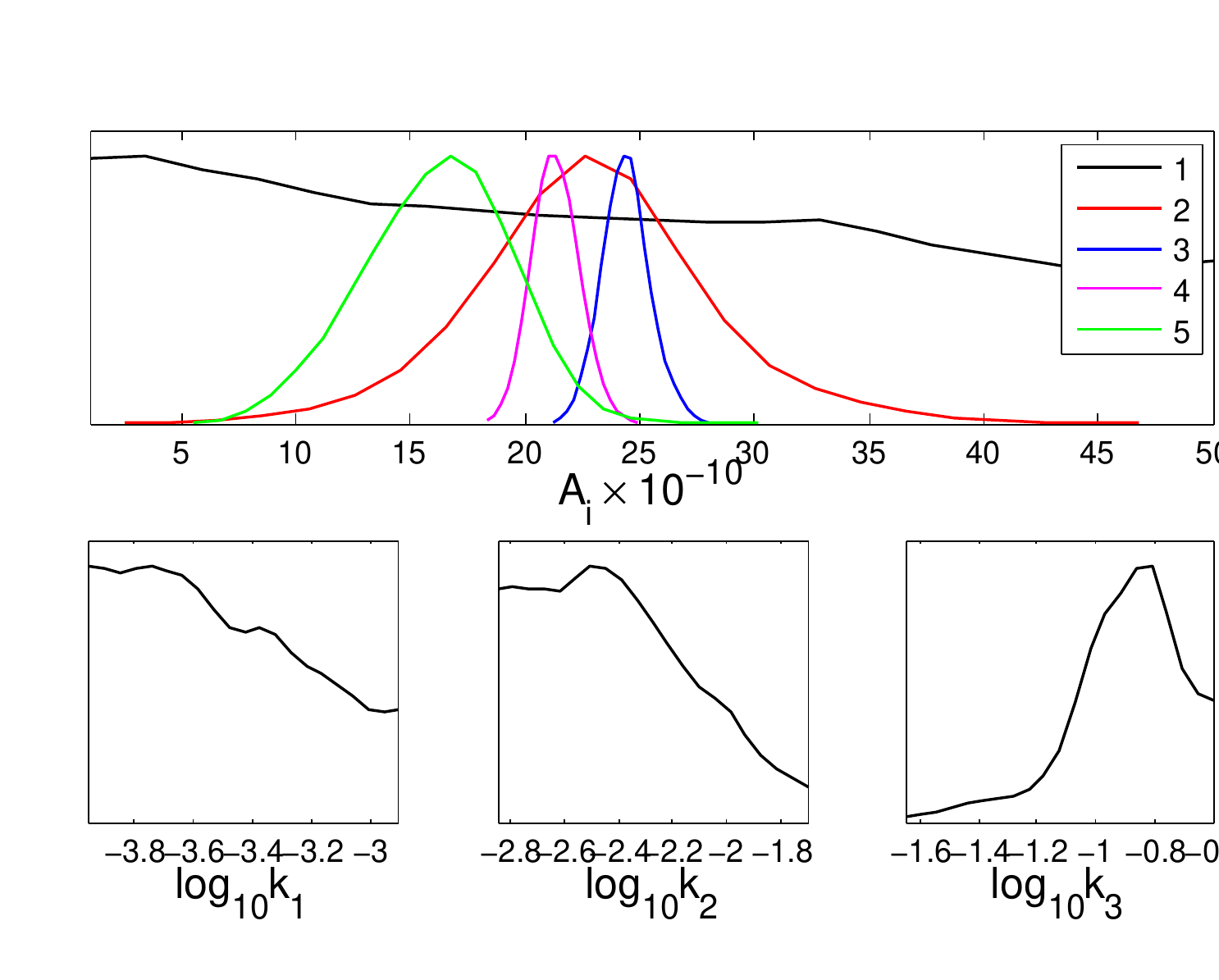} 
\end{array}$
\end{center}
\caption{Reconstruction of the primordial spectrum letting one  internal $k$-node (top) and two internal $k$-nodes
(bottom)  move freely in both amplitude $A_i$ and position $k_i$. The  right  panel corresponds to the
 1D marginalised posterior distribution of the amplitudes  and $k$-node position  in  each reconstruction.
  The top label  in each  panel  denotes the associated Bayes factor with  respect
to the base model (HZ) shown in Figure \ref{fig:recons_1} (a). }
\label{fig:mov_k}
\end{figure}

\noindent
The internal nodes generalise the spectral running by allowing slight variations in the 
interpolated slopes between external nodes.
Figure \ref{fig:mov_k} illustrates the reconstruction of the shape of the primordial spectrum 
from the mean posterior estimates - with $1\sigma$ error bars on the amplitudes - (left), along with 
the 1D marginalised posterior distributions on the parameters used  to describe the  spectrum (right).
 On large scales, the reconstructed shape of the one-internal-node model  ($k_1$) resembles a similar  
spectrum  to that obtained in Figure \ref{fig:recons_1} (c),
but now the probability distribution suggests a preferred turn-over  position localised at the largest scales. 
A  similar  turn-over  has also been  identified  using  principle
component  analysis \cite{Guo11}.
 In the two and three-internal-nodes cases, it is interesting to note that at small scales 
 the marginalised posterior peaks at scales where the combined
WMAP/ACT constraints are improved; at these scales ($0.1 < k \, [
  {\rm Mpc}^{-1}] < 0.14$) CMB data now considerably overlap with
measurements from SDSS DR7 LRG. This  reduced power at small scales
might be identified as a feature produced  from a phase transition in the  early universe \cite{Barriga01}. 
Both, the two and three-internal models (middle and bottom panel of Figure
\ref{fig:mov_k}), present a similar behaviour on the reconstructed spectra, also
seen on the marginalised posterior distributions.

\subsubsection{Cut-off and Broken spectra}

For completeness, we consider the possible  existence  of a large-scale cut-off  on the primordial
spectrum. 
A possible motivation to consider this  model has been discussed for  instance  by \cite{Contaldi03}.
In order to perform the reconstruction  for this particular 
 case we fix an extremal node at $k_{\rm max}$ with varying amplitude  $A_{{\rm s},k_{\rm max}}$
and let the cut-off scale $k_c$ vary across the prior $[k_{\rm min},k_{\rm max}]$ as well
as its amplitude $A_{{\rm s}, c}$.
The form of the spectrum is described as follows:
\begin{eqnarray} 
\mathcal{P_R}(k) = \left\{ \begin{array}{ll} 
0 						& \quad	k\le k_c\\ 
A_{{\rm s}, c}			      	& \quad	k_c< k< k_{\rm max} \\ 
A_{\rm s,max}	 			&\quad 	k\ge k_{\rm max} 
\end{array} \right. &&
\\ \nonumber \\
{\rm and\,\, with\,\,  linear\,\,interpolation\,\, for \quad}  k_c <  & k & < k_{\rm max}. \nonumber
\end{eqnarray}

\noindent
Also, we consider a broken spectrum which might have been produced
from phase transitions in the early universe. A similar broken
spectrum, motivated by double or multiple field inflation, has been
considered by \cite{Barriga01, Bridle03}.  This spectrum is obtained
by placing two nodes $k_1$ and $k_2$ within $[k_{\rm min}, k_{\rm max}
]$, and letting them move freely in amplitude $A_{{\rm s},k_i}$ and $k$-position, such that
\begin{eqnarray} 
\mathcal{P_R}(k) = \left\{ \begin{array}{ll} 

A_{{\rm s},k_1} 				& \quad	k<k_1\\ 
A_{{\rm s},k_2}			      	& \quad	k\ge k_2
\end{array} \right. && \\ \nonumber \\
{\rm and\,\, with\,\,  linear\,\,interpolation\,\, for \quad}  
 \quad& k_1& < k_2. \nonumber 
\end{eqnarray}

\begin{figure}
\begin{center}$
\begin{array}{cc}
 (k_c) \,\, $$\mathcal{B}_{k_c,1} =+2.98 \pm 0.30$$ &\\
 \includegraphics[trim = 1mm  2mm 5mm -5mm, clip, width=5cm, height=4.cm]{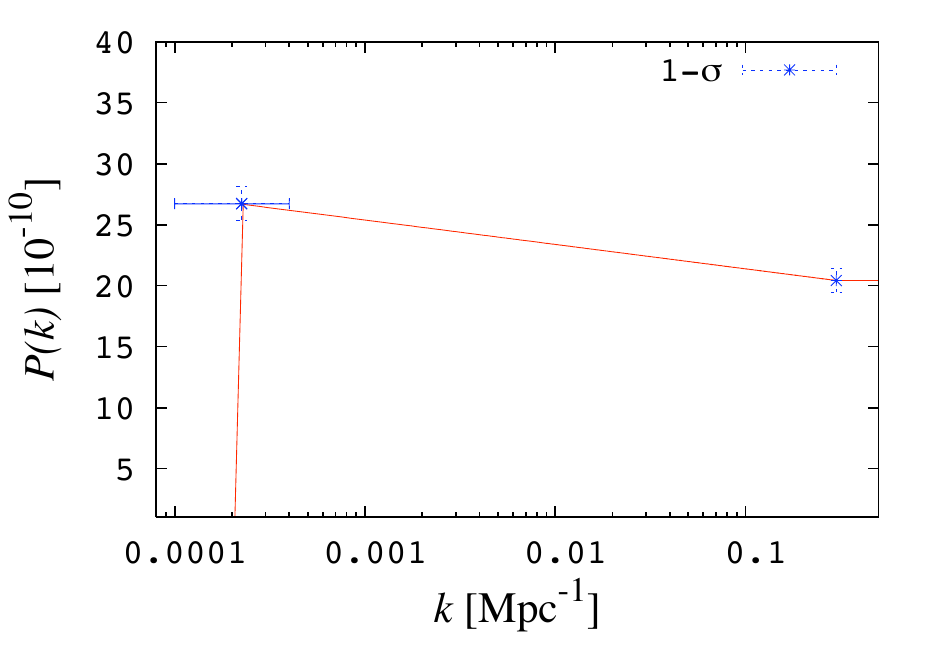}
\includegraphics[trim = 38mm  75mm 44mm 65mm, clip, width=5cm, height=4.cm]{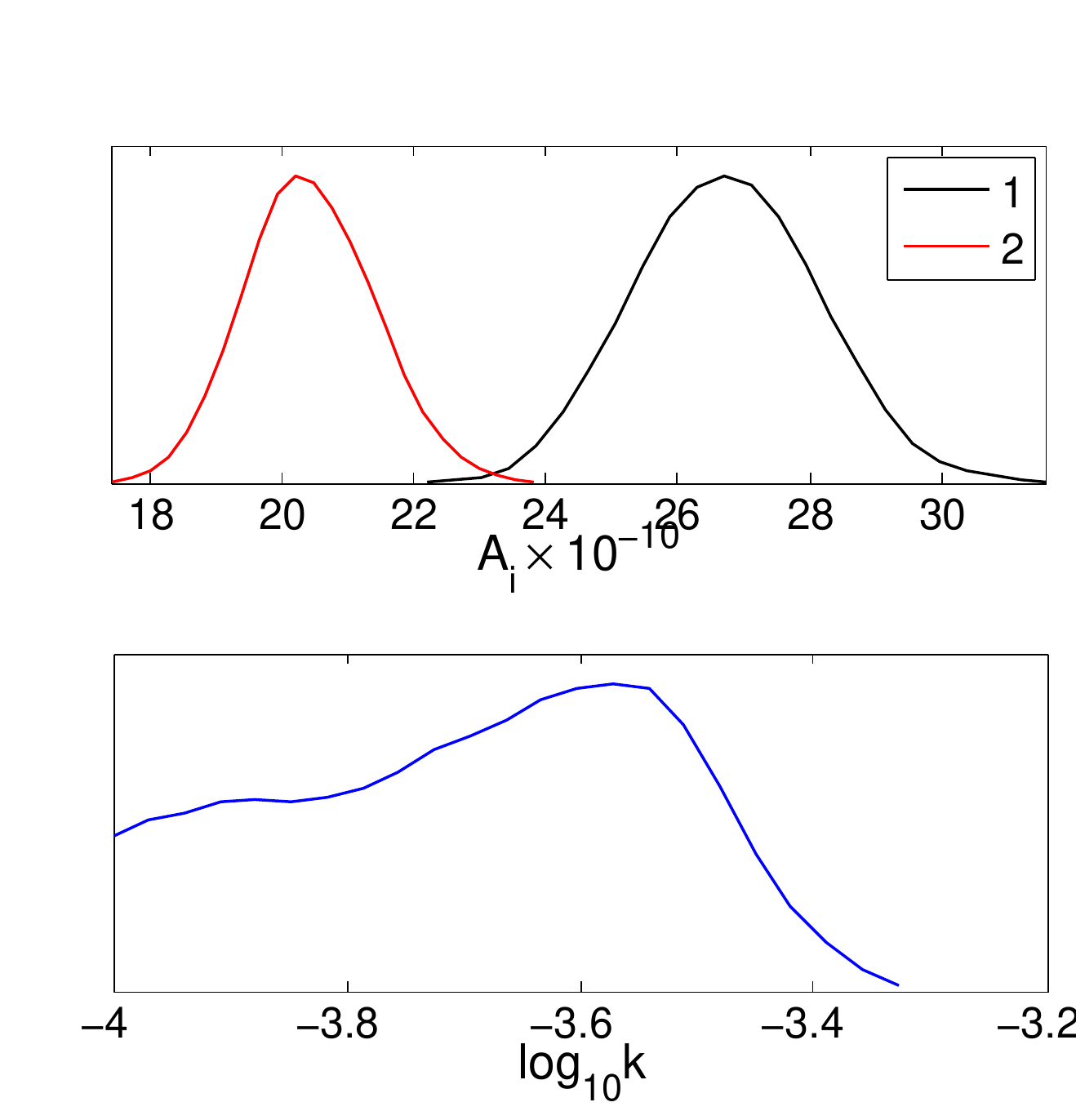} \\
 (k_b) \,\, $$\mathcal{B}_{k_b,1} =+ 2.38\pm 0.30$$ &\\
 \includegraphics[trim = 1mm  2mm 5mm -5mm, clip, width=5cm, height=4.cm]{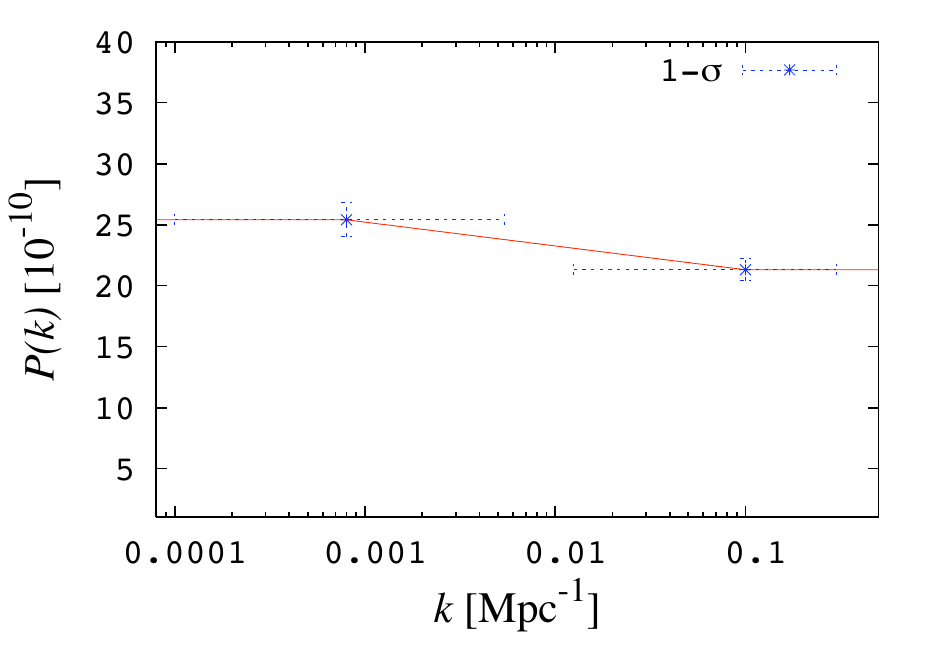}
\includegraphics[trim = 38mm  70mm 44mm 65mm, clip, width=5cm, height=4cm]{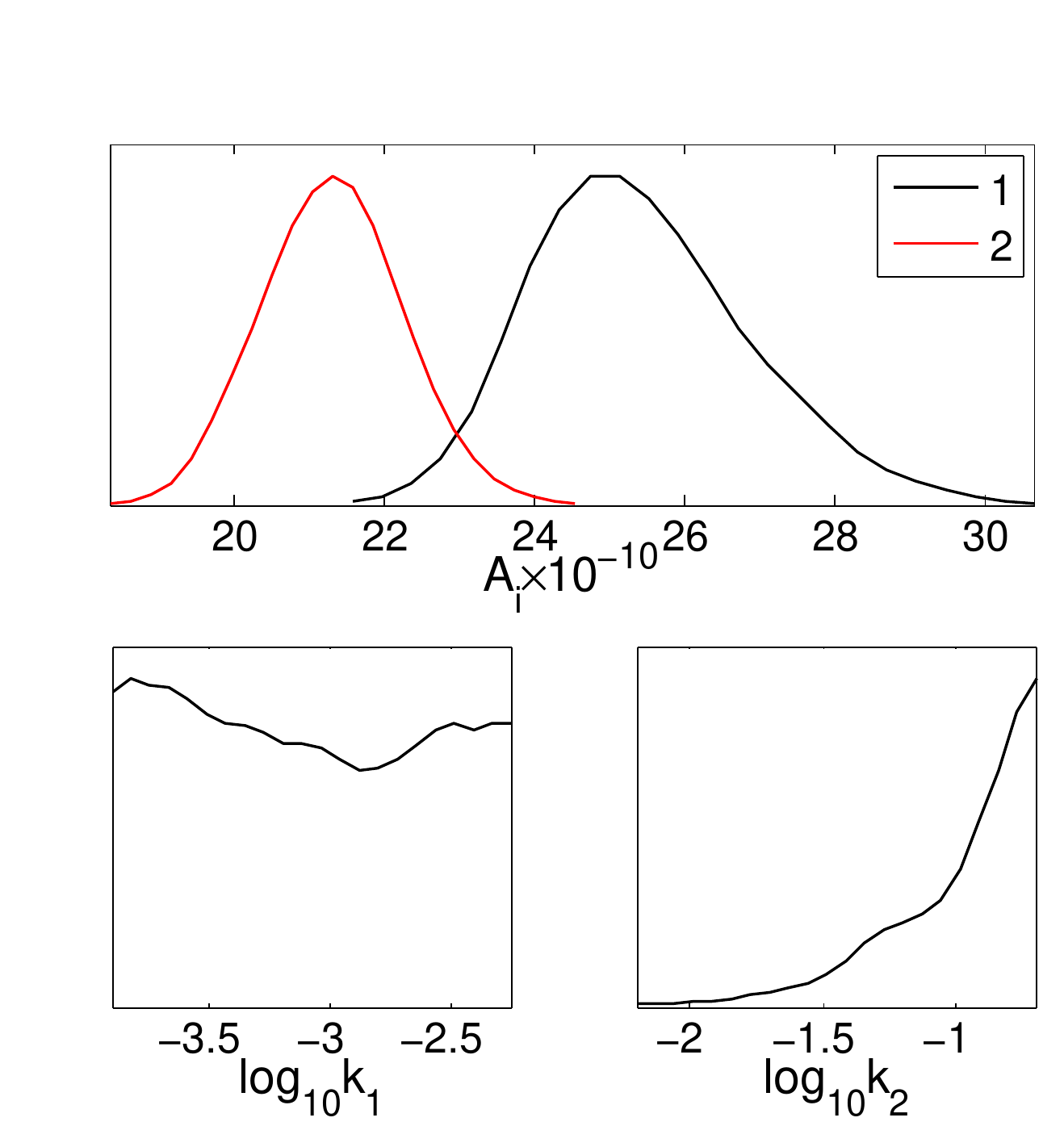} 
\end{array}$
\end{center}
\caption{Reconstruction of the large scale cut-off spectrum (top) and
broken spectrum (bottom); their corresponding  1D marginalised posterior distribution 
of the amplitudes $A_{\rm i}$ and node positions $k_i$  in  each  reconstruction (right).
  The top label  in each  panel  denotes the associated Bayes factor with  respect
to the base model (HZ) shown in Figure \ref{fig:recons_1} (a).}
\label{fig:cut}
\end{figure}

\noindent
 The  reconstruction of the cut-off and broken primordial spectra
 along  with 1$\sigma$ limits  of the marginalised  distributions are  shown in the left panels of Figure \ref{fig:cut}.
Their  corresponding posterior  distribution in  each parameter used to  describe the spectra are illustrated in the right panels.
The obtained best-fit parameters for the cut-off spectrum (top), show a preferred scale
at which the  power  drops  to  zero with  an upper limit  $\log_{10}k_c <-3.45$ at 95\% C.L. 
 Our  constraints  on $k_c$ also show  a significant likelihood at large scales, thereby disfavouring the presence
 of an abrupt cut-off. 
With respect to the broken model (bottom), on the other hand, the best fit parameters indeed predict a break in the 
primordial spectrum, located approximately at $\log_{10}k \simeq -2.2$.
 That could be  an indicative  of  the existence of a phase transition, and it is similarly obtained in 
 the two and three-internal-nodes models shown in  Figure \ref{fig:mov_k}. 
\\

In this section we have considered three types of spectra with different features:
turn-over, large scale cut-off and broken spectrum (Figures \ref{fig:mov_k}-\ref{fig:cut}). 
In each figure we have included the Bayes factor compared to the base model (HZ). 
According to the Jeffreys guideline the one-internal-node spectrum, shown in the  top panel of Figure \ref{fig:mov_k},
is significantly preferred over the cut-off and broken spectrum models,  $\mathcal{B}_{k_1,k_c} =+1.28 \pm 0.30$
and  $\mathcal{B}_{k_1,k_b} =+1.88 \pm 0.30$ respectively. 
Even  though the model with one-internal node  is described  by four parameters, when it is compared to the 
Harrison-Zel'dovich spectrum (with only one parameter), the Bayes factor  $\mathcal{B}_{k_1,1} =+4.26 \pm 0.30$
 shows that the presence of a turn-over is strongly favoured by current cosmological information and
significantly so when compared to the tilted spectrum (with  two parameters) 
$\mathcal{B}_{k_1,2} =+1.33 \pm 0.30$, see  Figure \ref{fig:recons_1}. 
Therefore, the presence of a turn-over in $\mathcal{P_R}(k)$ plays an  important  role in explaining current observations.
Notice that,  in the bottom panel of Figure
\ref{fig:mov_k}  the Bayesian evidence has dropped off, 
hence the reason  we have stopped the addition of nodes in the reconstruction process.

\subsection{Power-law and running spectra}

We have considered, so far, a $\mathcal{P_R}(k)$ shape reconstructed directly from data.
For comparison we include the standard approach by assuming a
power-law parameterisation in terms of a spectral amplitude
$A_{\rm s}$ and a spectral index or tilt parameter $n_{\rm s}$:

\begin{equation}
 \mathcal{P_R}(k) = A_{\rm s} \left( \frac{k}{k_0} \right)^{n_{\rm s}-1},
\end{equation}

\noindent
where the spectral index is expected to be close to unity; $k_0$
denotes the scale at which the tilted spectrum pivots, fixed to
$k_0=0.002$ ${\rm Mpc}^{-1}$. We again assume the prior $A_{{\rm
    s}}=[1,50] \times 10^{-10}$ on the amplitude, together with the
conservative prior $n_{\rm s} = [0.7,1.2]$ on the spectral index.
 We find a mean value of $n_{\rm s} =0.963\pm0.011$ which confirms that our constraints are 
 in good agreement with results from \cite{WMAP,ACT,SPT}.
As a further extension we consider possible deviations from power-law by allowing the spectral index to vary
as a function of scale   $n_{\rm s}(k)$. Then the primordial spectrum becomes 

\begin{equation}
 \mathcal{P_R}(k) = A_{\rm s} \left( \frac{k}{k_0} \right)^{n_{\rm s}-1+\frac{1}{2}\ln \left(\frac{k}{k_0}\right)n_{\rm run}},
\end{equation}

\begin{figure}[t!]
\begin{center}$
\begin{array}{cc}
(n_{\rm s}) \,\, $$\mathcal{B}_{n_{\rm s},1} =+3.25\pm 0.30$$  &\\
 \includegraphics[trim = 1mm  -2mm 5mm -5mm, clip, width=5cm, height=4cm]{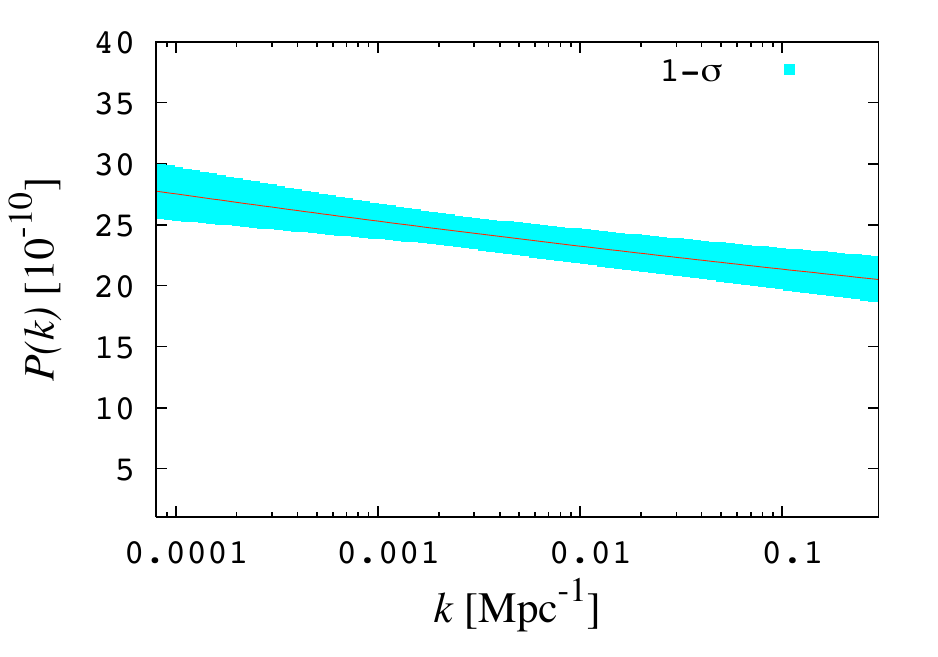}
\includegraphics[trim = 30mm  65mm 43mm 70mm, clip, width=5cm, height=4cm]{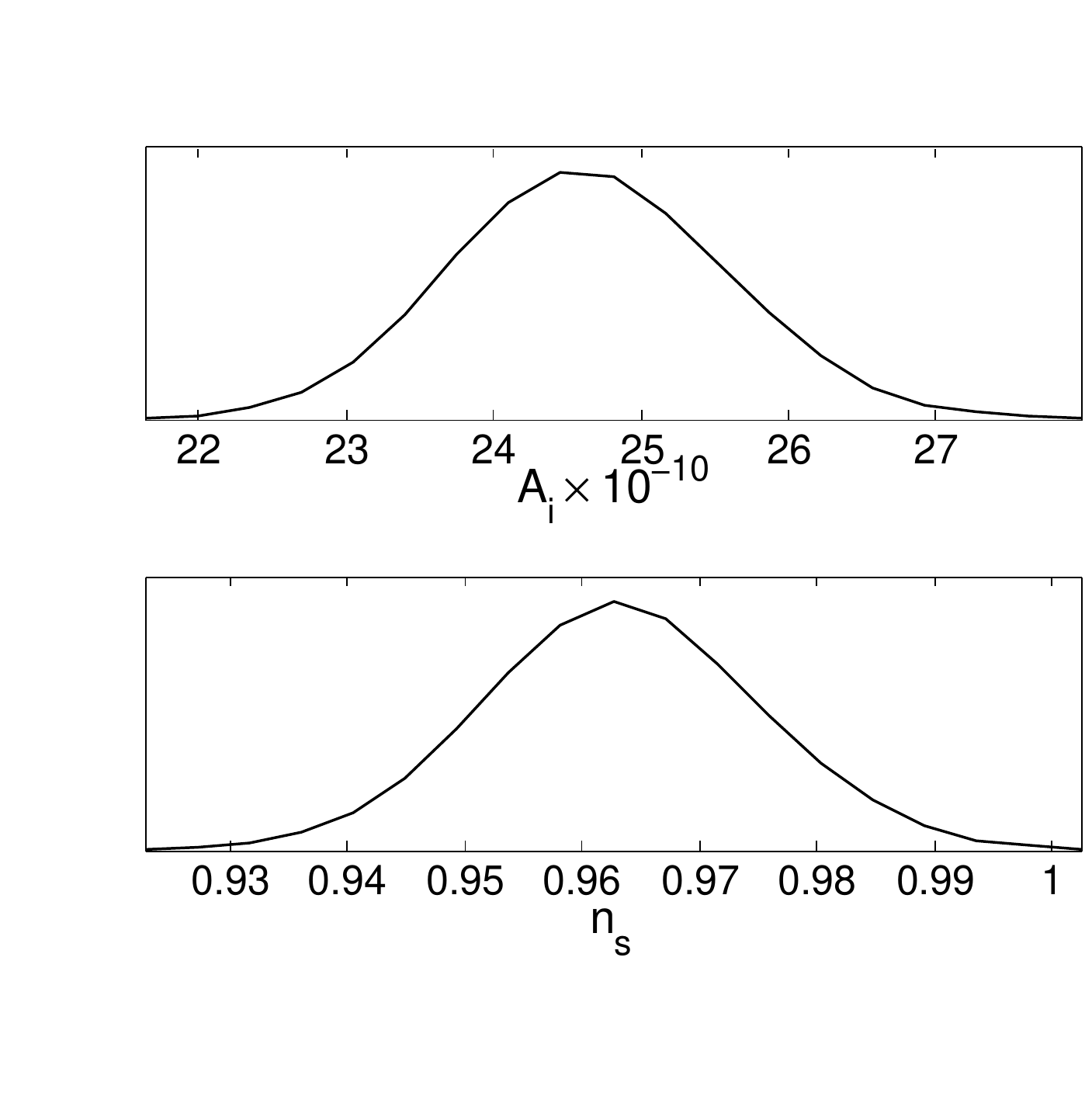} \\
(n_{\rm run}) \,\, $$\mathcal{B}_{n_{\rm run},1} =+2.06\pm 0.30$$&\\
 \includegraphics[trim = 1mm  -2mm 5mm -5mm, clip, width=5cm, height=4cm]{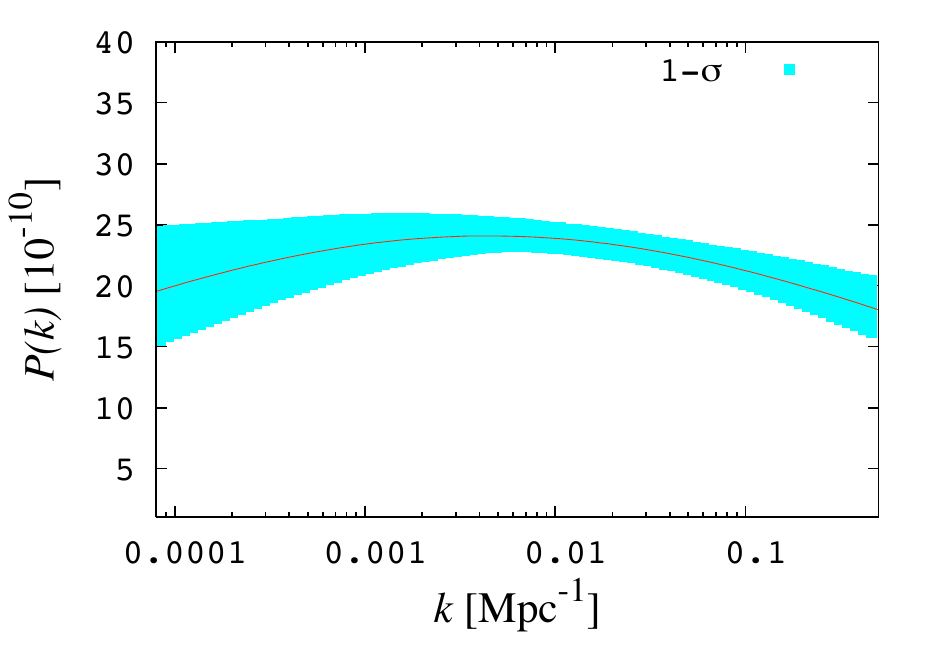}
\includegraphics[trim = 30mm  68mm 45mm 70mm, clip, width=5cm, height=4cm]{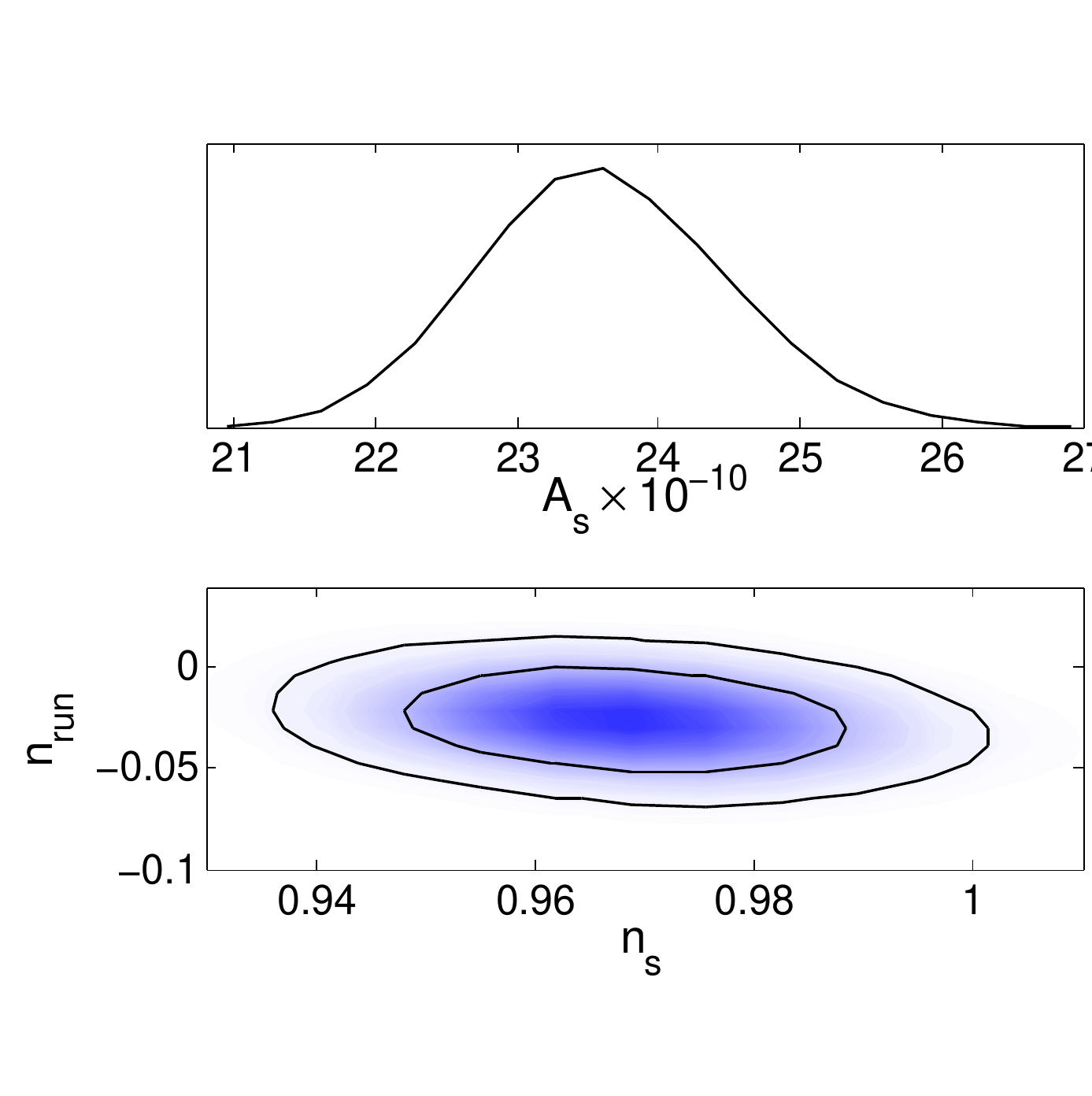} \\
\end{array}$
\end{center}
\caption{ Reconstruction of the primordial spectrum assuming  a simple  tilted parameterisation  with 
$n_{\rm s}$ (top) and including the running parameter $n_{\rm run}$ (bottom);
 the coloured  region  denotes $1\sigma$  error bands on the  reconstruction.
 Right: marginalised  1D and  2D probability posterior distributions for the  power spectrum parameters;
 2D constraints are plotted  with $1\sigma$ and $2\sigma$ confidence contours.
 The top label  in each  panel  denotes the associated Bayes factor with  respect
to the base model (HZ) shown in Figure \ref{fig:recons_1} (a). }
\label{fig:pow}
\end{figure}

\noindent
where $n_{\rm run}$ is termed the running parameter and is expected to be $n_{\rm run} \approx 0$ 
for standard inflationary models. In order to minimise the  correlation between $n_{\rm s}$ and $n_{\rm run}$
we have considered a pivot scale of  $k_0=0.015$ ${\rm Mpc}^{-1}$, 
 as pointed out by \cite{Cortes07}.
We use the same priors as above on $A_{\rm s}$ and $n_{\rm s}$, and
the conservative prior $n_{\rm run} = [-0.3,0.3]$ on the running parameter.
From the combined dataset we  find the marginalised posteriors  show a preference for a 
negative running parameter $n_{\rm run}=-0.026\pm0.015$  and $n_{\rm s}=0.968\pm0.011$ for the spectral
index, 
as  expected by \cite{WMAP,ACT,SPT}.
%
Figure \ref{fig:pow}   shows  the marginalised posterior distributions for the parameters used
to describe $\mathcal{P_R}(k)$ and the obtained spectrum from mean  posterior  estimates of 
a simple  tilted parameterisation with $n_{\rm s}$ (left)  and including the running parameter 
$n_{\rm run}$ (right) respectively. In each panel we have included the Bayes factor compared to the base model (HZ).
\\

According to the Jeffreys guideline, present observations significantly prefer a simple tilted 
model  when  compared  to a model which includes a running parameter by a factor 
$ \mathcal{B}_{n_{\rm s},n_{\rm run}} =+1.19\pm 0.30$. Similarly, a tilted  spectrum is strongly 
preferred when compared to the HZ model: $\mathcal{B}_{n_{\rm s},1} =+3.25\pm 0.30$.
We also confirm the agreement between the simple tilted model and the 
two-fixed-noded spectrum through its Bayes factor, shown in Figure \ref{fig:recons_1} (b).   
An important point to emphasise is that the simple tilt and running model present a significantly 
and  strongly disfavoured
Bayes factor, $\mathcal{B}_{n_{\rm s}, k_1} =-1.01\pm 0.30$,
 $\mathcal{B}_{n_{\rm run},k_1 } =-2.20\pm 0.30$,  compared  to the  reconstructed 
one-internal-node spectrum shown in Figure \ref{fig:mov_k}. Thus a simple  power-law  parameterisation
seems to be not  enough to describe current cosmological observations, hence slight deviations  of it  should
be  taken into account.

\subsection{Modified Power-law spectrum}
 
 We have observed that models which present a turn-over 
 at large scales are slightly preferred by the evidence.
 Based on this observation, we suggest the following phenomenological shape for the primordial power spectrum:
 
 \begin{equation}
 \mathcal{P_R}(k)=  A_{\rm s} \frac{ \left(\frac{k}{k_{\rm v}} \right)^{n_{\rm v}}}{ \frac{k}{k_{\rm v}}+1}.
 \end{equation}
 
 \noindent
 In this particular parameterisation,  assuming 
 $n_{\rm v}<1$, the parameter $k_{\rm v}$ determines the transition between 
  a standard power-law model with red tilt ($k\gg k_{\rm v}$) to a blue tilt model ($k \ll k_{\rm v}$):

\begin{equation}
 \mathcal{P_R}(k) =  A_{\rm s}  \left\{
  \begin{array}{l l}
     \left( \frac{k}{k_{\rm v}} \right)^{n_{\rm v}-1} & \quad k \gg k_{\rm v}, \\
     \left( \frac{k}{k_{\rm v}} \right)^{n_{\rm v}} & \quad k \ll k_{\rm v},
  \end{array} \right.
\end{equation}

\begin{figure}[t!]
\begin{center}
$\begin{array}{cc}
(n_{\rm v}) \,\, $$\mathcal{B}_{n_{\rm v},1} =+4.65\pm 0.30$$ \\
 \includegraphics[trim = 1mm  -2mm 5mm -5mm, clip, width=5cm, height=4.5cm]{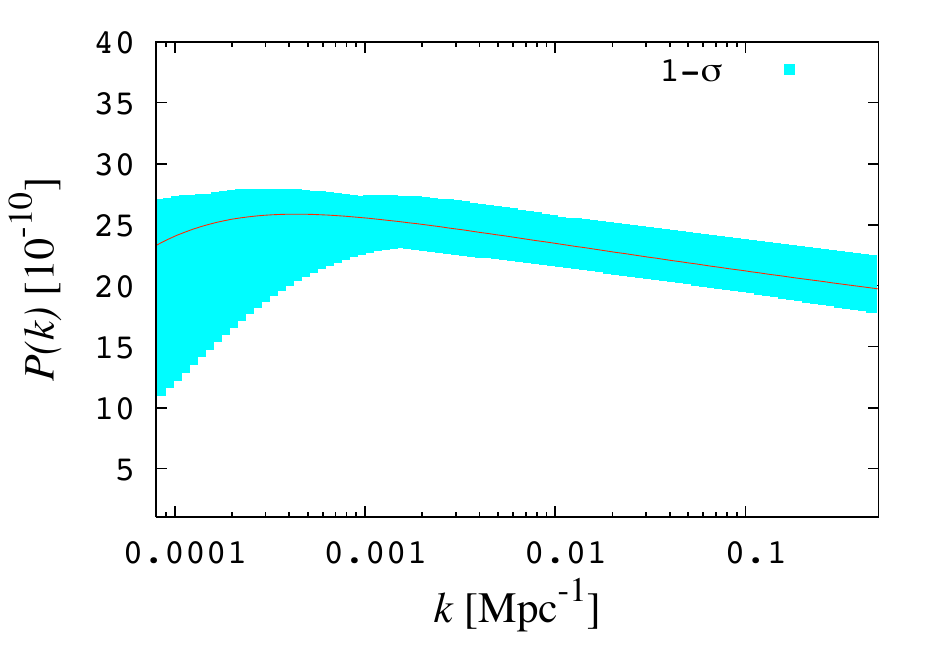}
\includegraphics[trim = 30mm  62mm 45mm 65mm, clip, width=5cm, height=4.5cm]{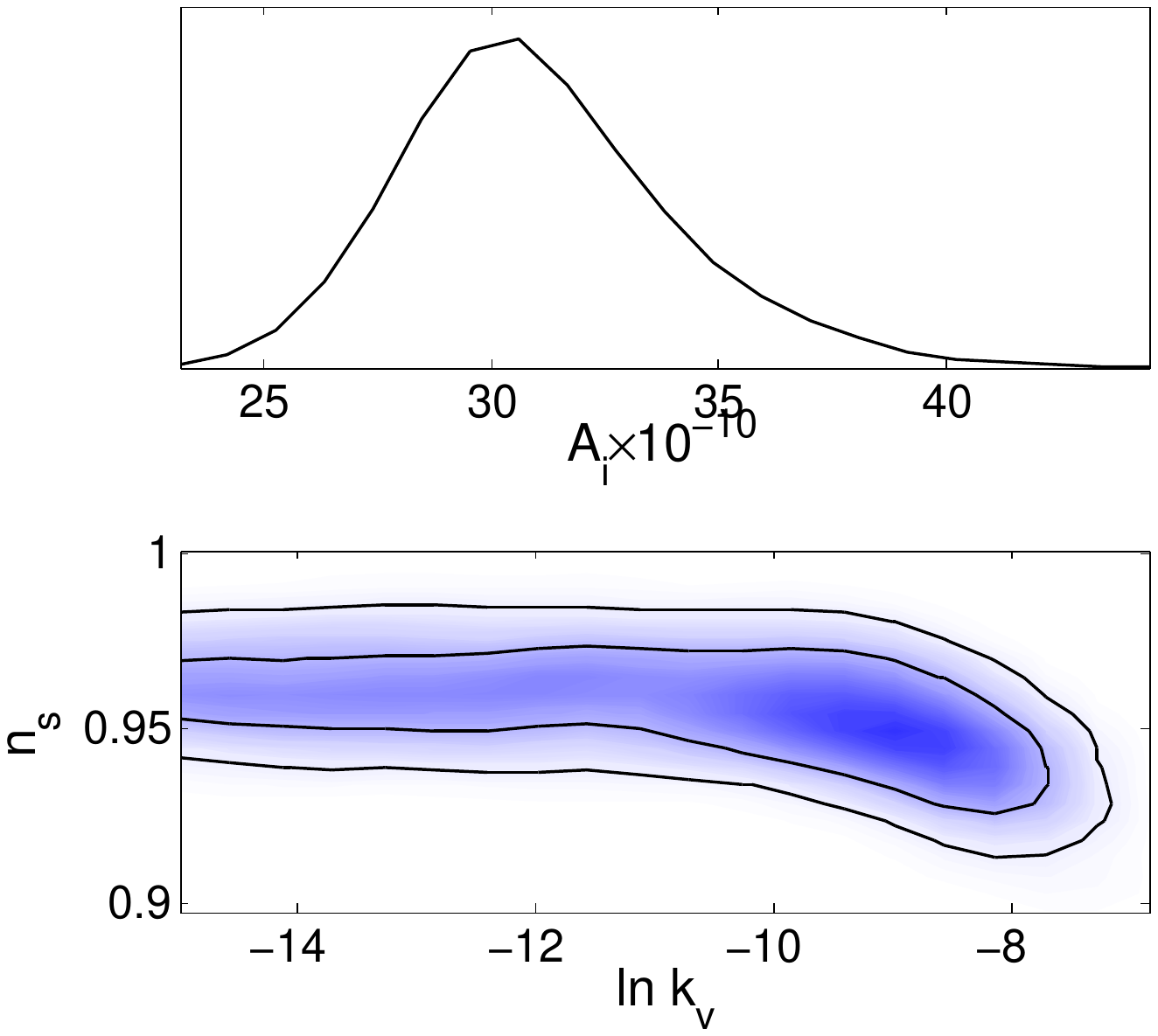} 
\end{array}$
\end{center}
\caption{ Reconstruction of the primordial spectrum  from mean posterior estimates  of the  modified
power-law parameterisation, along  with $1\sigma$ error bands (left).
 Right: marginalised  1D and  2D probability posterior distributions for the  power spectrum parameters;
 2D constraints are plotted  with $1\sigma$ and $2\sigma$ confidence contours.
 The top label  in each  panel  denotes the associated Bayes factor with  respect
to the base model (HZ) shown in Figure \ref{fig:recons_1} (a). }
\label{fig:mod_ns}
\end{figure}

\noindent
where the prior on $n_{\rm v}$ is similarly chosen to the spectral index $n_{\rm s}$ in the power-law
parameterisation: $n_{\rm v}=[0.7,1.2]$. We expect the constraints on the parameter $k_{\rm v}$ are mainly
located on large scales, hence, for this extra-parameter we consider the following flat  prior $\ln k_{\rm v}=[-15,-5]$.
The reconstruction of the  shape  of this spectrum along with the posterior distribution
in each parameter are shown  in Figure \ref{fig:mod_ns}.
We observe the constraints for the new tilt-parameter $n_{\rm
  v}=0.955\pm0.014$ are similar to those obtained from the power-law
models, where the scale-invariant spectrum is ruled out at a high
confidence level and the spectrum exhibits a red tilt at small scales.
The marginalised posterior probability on $k_{\rm v}$ shows the
existence of a blue-tilted spectrum on large scales $\ln
k_{\rm v}<-8.1$ at 95\% C.L.  Hence, the global shape for this
spectrum presents a slight running behaviour with reduced power at both
large and small scales, compared to the simple tilt parameterisation.
\\

The modified power-law parameterisation decisively rules out the HZ,
since the Bayes factor between the models is $\mathcal{B}_{n_{\rm v},1}
=+4.65\pm 0.30$.  Moreover, even though the modified
power-law model has an extra parameter compared to the simple
power-law model, the data significantly prefer it with $\mathcal{B}_{n_{\rm
    v},n_{\rm s} } =+1.40\pm 0.30$.

\subsection{Lasenby \&  Doran spectrum}

\begin{figure}[t!]
\begin{center}
$\begin{array}{cc}
(n_{\rm LD}) \,\, $$\mathcal{B}_{{\rm LD},1} =+4.94\pm 0.30$$ \\
 \includegraphics[trim = 1mm  -2mm 5mm -5mm, clip, width=5cm, height=4.5cm]{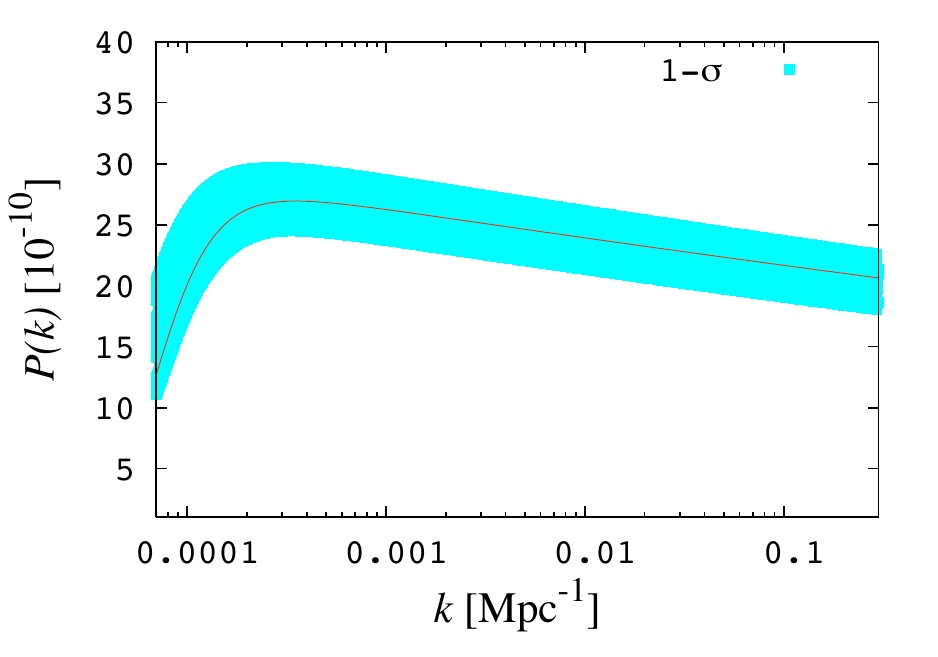}
\includegraphics[trim = 28mm  63mm 48mm 65mm, clip, width=5cm, height=4.5cm]{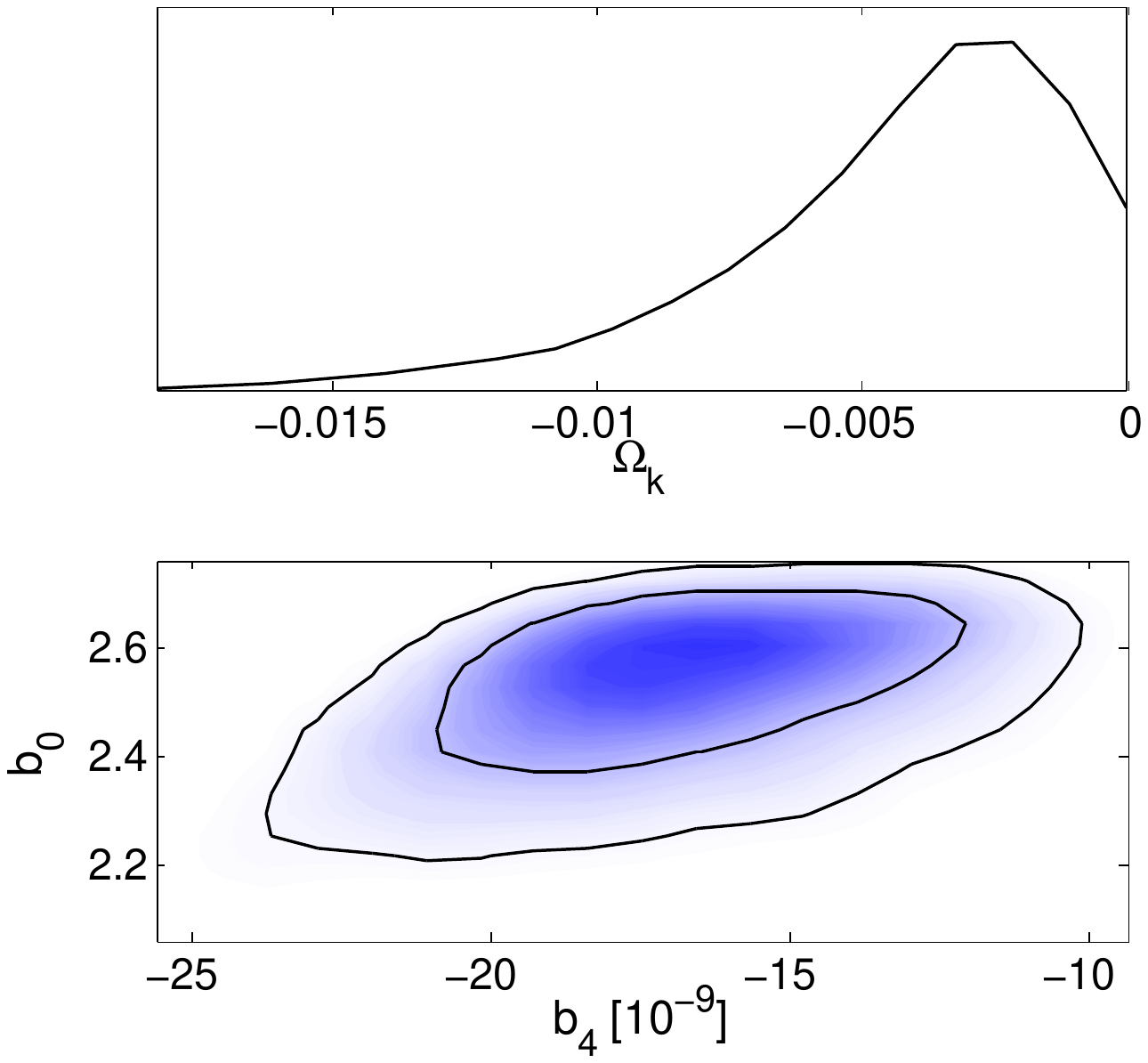} 
\end{array}$
\end{center}
\caption{Reconstruction of the primordial spectrum assuming  a Lasenby  \&  Doran model 
along  with $1\sigma$ error bands (left).
 Right: marginalised  1D and  2D probability posterior distributions for the  power spectrum parameters;
 2D constraints are plotted  with $1\sigma$ and $2\sigma$ confidence contours.  
 The top label  on the  panel  denotes the associated Bayes factor with  respect
to the base model (HZ) shown in Figure \ref{fig:recons_1} (a). }
\label{fig:LD}
\end{figure}

The Lasenby and Doran model (LD, \cite{Lasenby05}) is based on the
restriction of the total conformal time available in the entire
history of a closed Universe.  
The primordial  spectrum  derived  from this model naturally incorporates an exponential cut-off on large scales 
which might provide  a possible explanation for the lower-than-expected CMB power spectrum
at  low multipoles. 
On small scales, the relationship between $\mathcal{P}^{1/2}_\mathcal{R}(k)$ and $\ln k$ is linear, thus predicting a reduced 
power at large $k$ as compared to a simple tilted spectrum (for which $\ln \mathcal{P}^{1/2}_\mathcal{R}(k)$ 
versus $\ln k$ is linear).
For further details about the LD model see, for instance \cite{Lasenby03, Lasenby04}.  
In this model, the resultant primordial
spectrum from the inflationary phase depends upon just two parameters
$\{b_0,b_4\}$. The parameter $b_0$ is mainly restricted by the number  of $e$-folds
$N \approx 2\pi b_0^2\approx50$,
whereas $b_4$ controls the initial curvature and must be negative and such that the dimensionless 
combination $|b_4|\mu^{-4/3}$ is of order unity. 
Here $\mu$ is the mass of the scalar field, and this along with $b_0$ and $b_4$ control the magnitude of 
the inflaton field, and how long the inflationary period lasts.
 To compute the LD spectrum we refer our analysis
to \cite{Vazquez11}. We have also chosen the priors based on the same
paper: $\Omega_k =[-0.05, 10^{-4}]$, $b_0 =[1,4] $,  $b_4=[-30,-1]\times 10^{-9}$. 
Figure \ref{fig:LD} shows the reconstructed
shape of the primordial spectrum along with the posterior distribution in each additional parameter for this model;
 the constraints on the present hubble parameter are  $H_0=69.4 \pm 1.4$,
 whereas the number of $e$-folds is $N= 50.6\pm4.3$.     
From the top  label of Figure \ref{fig:LD}, we observe the LD model
is significantly preferred over the simple power-law parameterisation
with a Bayes factor of $\mathcal{B}_{{\rm LD}, n_{\rm s} } =+1.69\pm
0.30$ and decisive when compared  to the HZ  spectrcum:  $\mathcal{B}_{{\rm LD},1} =+4.94\pm 0.30$.
\\

\begin{figure}[t!]
\begin{center}
$\begin{array}{ccc}
 \includegraphics[trim = 1mm  -2mm 1mm -10mm, clip, width=6.5cm, height=4.5cm]{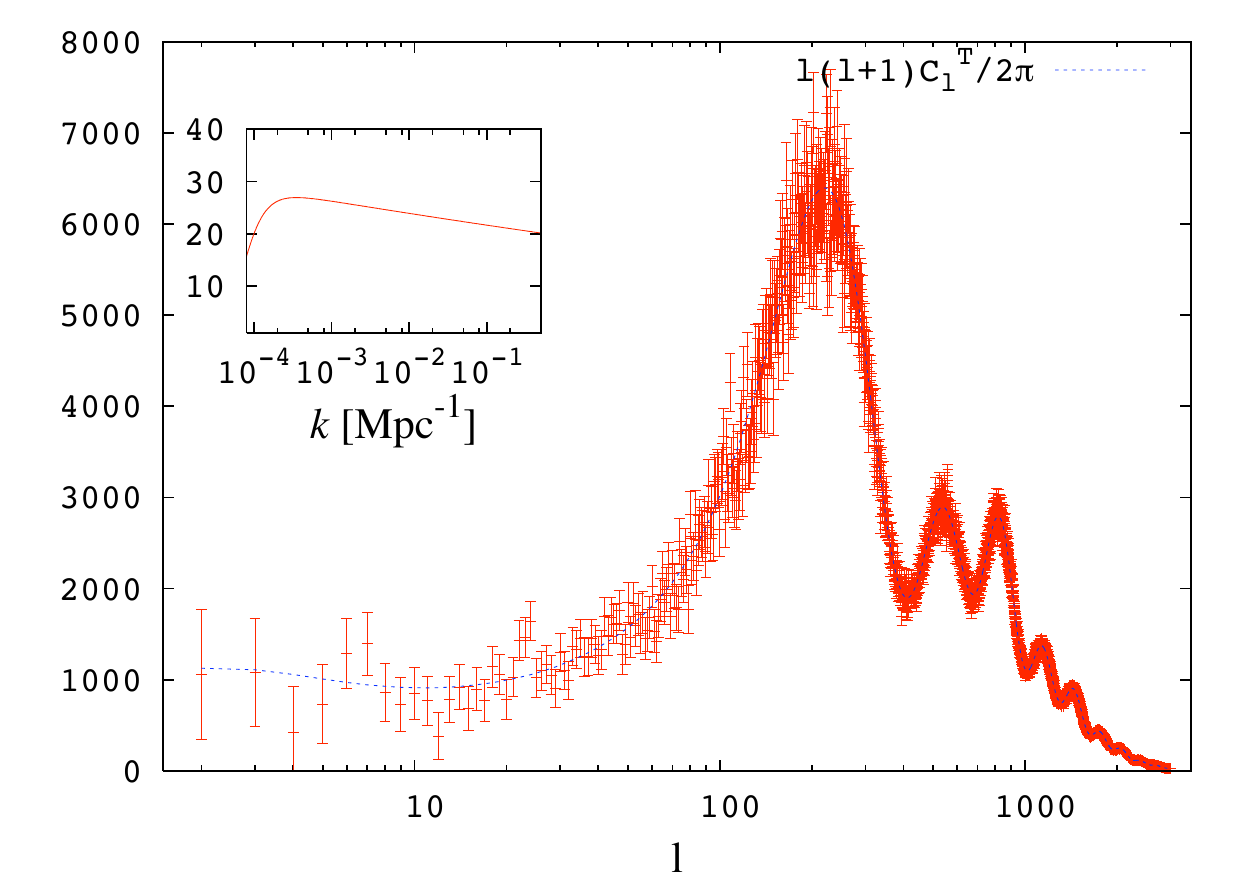}
\quad 
\quad
 \includegraphics[trim = 1mm  1mm 5mm -6mm, clip, width=5.5cm, height=4.5cm]{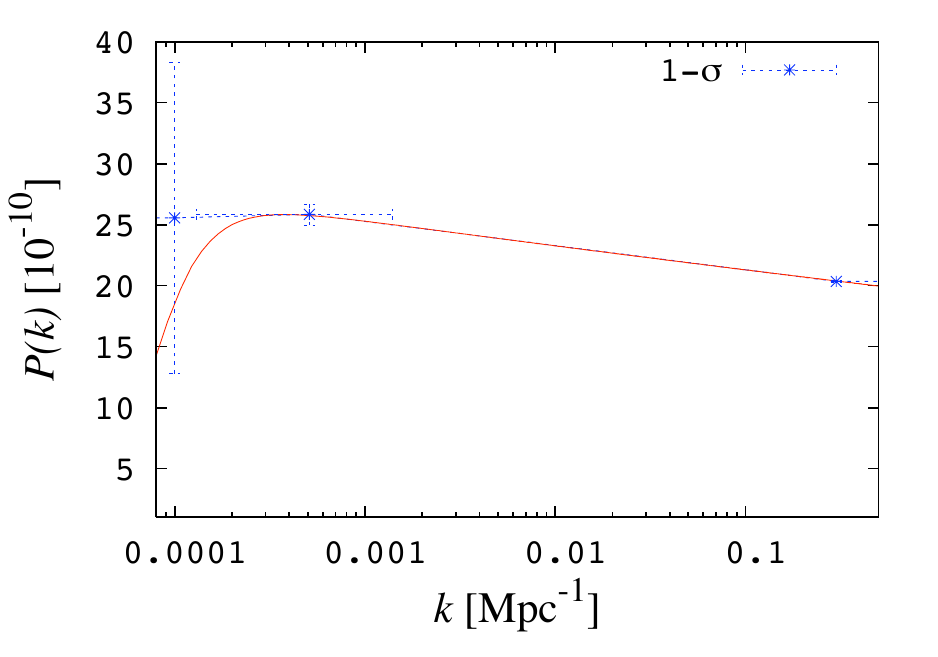}
\end{array}$
\end{center}
\caption{Reconstruction of the Lasenby  \&  Doran primordial spectrum  based  on the binning  format
with linear interpolation described
in Section \ref{sec;k_i}. We have  assumed  an idealised  CMB spectrum with limitation only  due to cosmic variance (left).
Right panel shows  the  reconstructed spectrum in the  binning  format  together  with  the LD input spectrum
(solid  line). }
\label{fig:LD_noise}
\end{figure}

Indeed, the LD model has the largest evidence of all the models
investigated, followed closely by the modified power-law spectrum.  It
should be noted, however, that the latter was constructed specifically
to exhibit a turn-over on large scales, having already found that the
data prefer such a feature. A fairer comparison would be between the
LD model and the third most favoured model, namely the
one-internal-node linear-interpolation model described in Section
\ref{sec;k_i}, since both of these models were proposed {\em a
  priori}. 
To this end, and as a check on our analysis, we use the
best-fit LD model (shown in Figure \ref{fig:LD}) as the input spectrum
to simulate an idealised CMB observation containing only
cosmic-variance-limited noise.  Figure \ref{fig:LD_noise} (left panel)
shows the resulting CMB temperature spectrum.  We then analysed these
simulated data using the one-internal-node linear interpolation model
to reconstruct the primordial power spectrum. Figure
\ref{fig:LD_noise} (right panel) shows the resulting reconstruction
(dotted line), which recovers well the shape of the input LD spectrum
(solid line), except on the very largest scales, where there is little
information in the simulated CMB data.
Moreover, the reconstructed
spectrum has a similar shape to the one obtained from real data using
the one-internal-node model (see Figure~\ref{fig:mov_k}). We may
therefore understand the higher evidence for the LD model spectrum as
resulting from its similar quality fit to the data, but requiring
fewer free parameters than the one-internal-node linear-interpolation
model.
\\

Finally, we note that, in the node-based reconstruction, the use of
linear interpolation between the nodes may seem crude. It is
straightforward to generalise the node-based approach to more
sophisticated interpolation schemes, but this may not always yield
better results. In Appendix \ref{sec:app}, we illustrate this point by
reanalysing the simulated CMB data using a cubic spline interpolation
through the nodes, thus allowing one to reconstruct a smooth shape for
the primordial spectrum, but one that is less satisfactory than that
obtained using linear interpolation.

\section{Discussion and Conclusions}
\label{sec:results}

In this paper we have attempted to fit an optimal degree of structure
for the primordial power spectrum of curvature perturbations using
Bayesian model selection as our discriminating criterion.  We have
modeled the spectrum as a linear interpolation between a set of `nodes'
with varying amplitude and $k$-position.  We have also explored
different parameterisations of the primordial spectrum which include:
a power-law parameterisation with both tilt and running parameter,
a modified power-law spectrum and the Lasenby \& Doran (LD) model.
\\

\begin{figure}[t!]
\begin{center}
\includegraphics[trim = 5mm  105mm 5mm 110mm, clip, width=15cm, height=3.7cm]{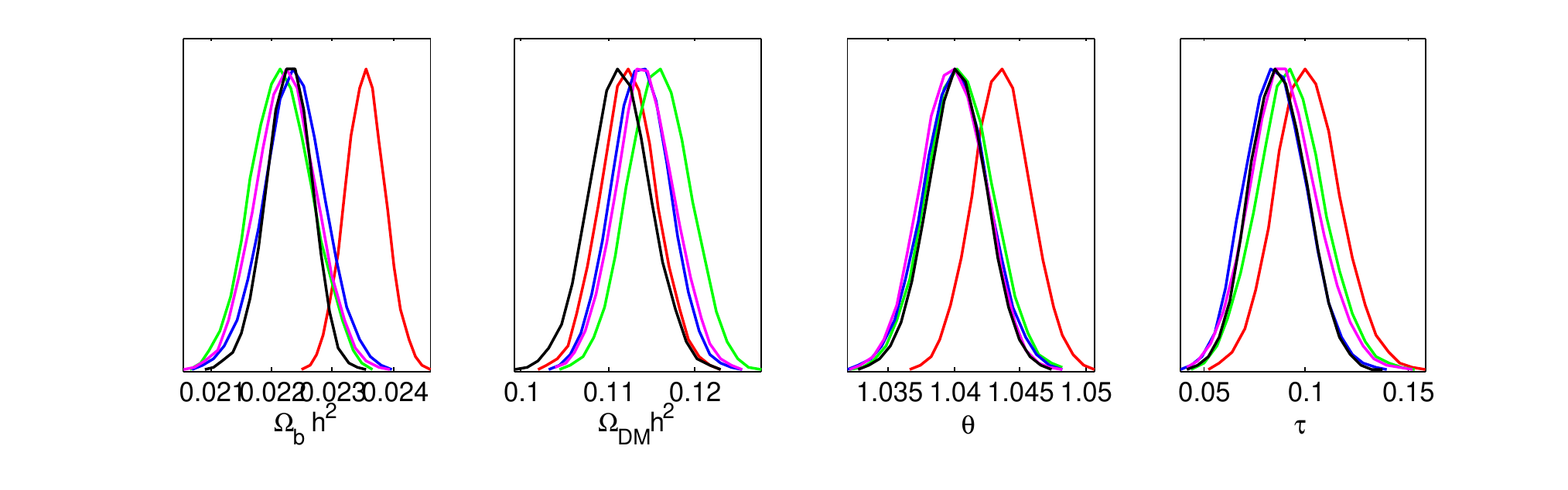} 
\includegraphics[trim = 3mm  105mm 3mm 110mm, clip, width=13cm, height=3.5cm]{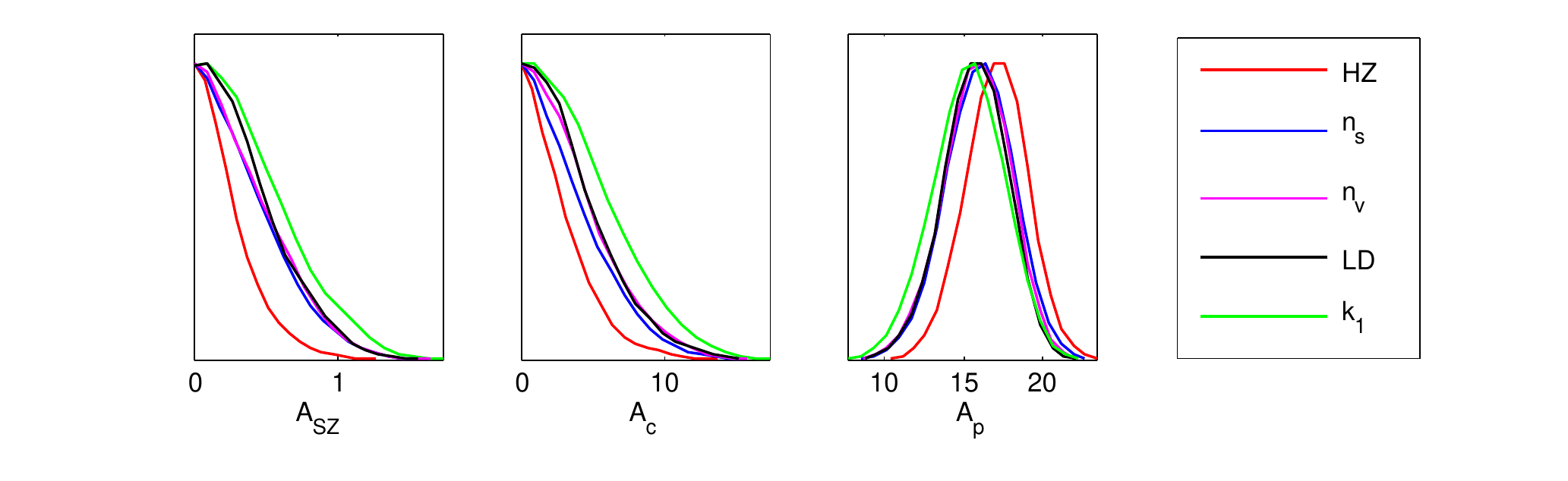} 
\end{center}
\caption{1D marginalised  posterior  distributions for the primary (top) and secondary (bottom)
 cosmological parameters, for each corresponding model  listed  on the bottom-right box.}
 \label{fig:pos_all}
\end{figure}

 All the considered models have in common the standard cosmological
 parameters: {$\Omega_{\rm b} h^2$, $\Omega_{\rm DM} h^2$, $\theta$,
   $\tau$}, as well as the secondary parameters: $A_{SZ}, A_p, A_c$.
 Thus, priors on these parameters remained the same in each model.  The
 best-fit values for these standard parameters are consistent with those
 obtained using the concordance 6-parameter model with power-law
 primordial spectrum.  We show, in Figure \ref{fig:pos_all}, 1D
 marginalised posterior distributions for the cosmological parameters
 of each of the preferred models.  We observe that the values of the
 standard parameters remain well constrained despite the addition of
 extra freedom on the shape of the primordial spectrum, although the
 constraints resulting from the HZ spectrum clearly differ from the
 others.  Note also that the constraints on the parameters corresponding
 to the LD model are slightly tighter than the rest of the models.
\\

We have considered wide-enough priors in our analysis, such that they
do not interfere with the inferred parameter values. We used priors on the
amplitudes of $A_i=[1,50]\times 10^{-10}$ and on spectral indices of $n_{\rm
  s}=n_{\rm v}=[0.7,1.2]$, while the parameters describing the
$k$-space have physical priors restricted by $[k_{\rm min}, k_{\rm
    max}]$. We also compute the Bayesian evidence for a wider prior
range of $n_{\rm s}=n_{\rm v}= [0.5,1.5]$ and
$b_4=[-50,-1]\times 10^{-9}$, to illustrate the robustness of a model
over small variations of the prior range: \\
\[ \mathcal{B}_{n_{\rm s}, 1 } =+2.25\pm 0.30 \qquad  ({\rm wide\, priors})\]
\[ \mathcal{B}_{n_{\rm v}, 1 } =+4.24\pm 0.30 \qquad  ({\rm wide\, priors})\]
\[ \mathcal{B}_{{\rm LD}, 1 } =+4.47\pm 0.30 \qquad  ({\rm wide\, priors})\]

\noindent
We observe that  even when wider  priors  are  considered, the  HZ model is strongly disfavoured
when compared  to  $n_{\rm v}$ and LD models. Similarly, the  simple  tilted  model is still  significantly
disfavoured.  
\\

\begin{figure}
\begin{center}
  \begin{minipage}[c]{0.43\textwidth}
  
\begin{tabular}{cccc} 
\cline{1-4}\noalign{\smallskip}
\hline
\vspace{0.2cm}
{\bf Model}&  N$_{\rm  par}$ &  -$2\Delta \ln \mathcal{L}_{\rm max}$& Bayes factor   
\\ \hline
\vspace{0.2cm}
HZ &			8		&0.0&	$\mathcal{B}_{1,1}$  	   	$=+0.0\pm 0.3$    	 	\\
\vspace{0.2cm}	
$n_{\rm s}$&	9		&-8.6&	$\mathcal{B}_{n_{\rm s},1}$	$=+3.3\pm0.3$				\\
\vspace{0.2cm}
$n_{\rm v}$&	10		&-9.4&	$\mathcal{B}_{n_{\rm v},1}$		$=+4.7\pm0.3$				\\
\vspace{0.2cm}
LD&			10		&-9.4&	$\mathcal{B}_{{\rm LD},1}$		 $=+4.9\pm0.3$				\\
\vspace{0.2cm}
 $k_1$&		11		&-9.1 &	$\mathcal{B}_{k_1,1}$			$=+4.3\pm 0.3$				\\
\hline
\hline
\end{tabular} 

  \end{minipage}
  \begin{minipage}[c]{0.53\textwidth}
 \includegraphics[trim = -30mm  -1mm 5mm -5mm, clip, width=8.5cm, height=5.5cm]{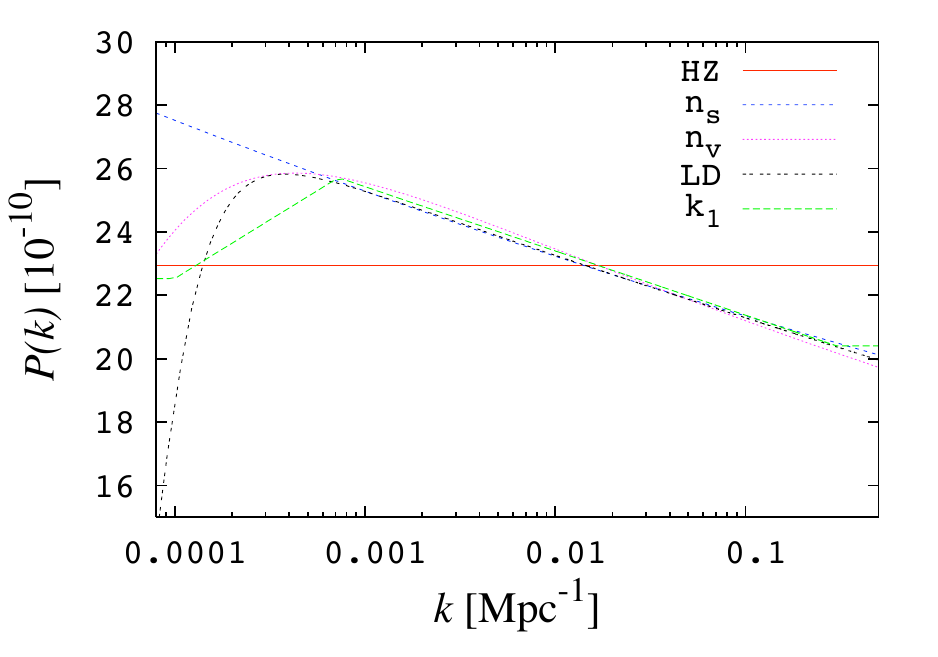}
  \end{minipage}
\end{center}
\caption{Comparison of the  primordial power spectra for the preferred models along with their Bayesian evidence.
We also  include the maximum likelihood $\mathcal{L}_{\rm max}$ for  a model with number of parameters N$_{\rm  par}$.  
Each Bayes factor is compared respect to the one-node model (HZ).  }
\label{fig:cosmo}
\end{figure}

To summarise  the analysis, in Figure \ref{fig:cosmo} we plot the reconstructed spectra for the preferred selected models
together with their corresponding Bayesian evidence.  
It shows that the HZ spectrum  is decisively excluded as  a  viable model
to  describe  $\mathcal{P_R}(k)$.
The preferred model  given current  observations  is provided  by the LD model followed by a 
modified power-law  version. We have found that the  power-law parameterisation, including
either  cases $n_{\rm s}$ and $n_{\rm s} + n_{\rm run}$, are both  significantly  disfavoured.
The presence of a turn-over at large  scales \footnote{At the largest  scales, the addition of tensor perturbations might 
considerably affect the shape of the primordial spectrum. Hence, we consider  this possibility in a more detailed future work. } 
and the reduced power at small scales seem to
provide  an important contribution on choosing the best-fit model
through its Bayesian evidence.
\\

\acknowledgments
This  work was carried out largely on the Cambridge High Performance 
Computing cluster,  {\sc DARWIN}. JAV is supported by CONACYT M\'exico.


\appendix

\section{Cubic spline interpolated spectrum} 
\label{sec:app}

\begin{figure}[h!]
\begin{center}
$\begin{array}{ccc}
 \includegraphics[trim = 1mm  0mm 1mm -1mm, clip, width=6.5cm, height=4.1cm]{Noise.pdf}
\quad 
\quad
 \includegraphics[trim = 1mm  2mm 5mm -1mm, clip, width=5.5cm, height=4.1cm]{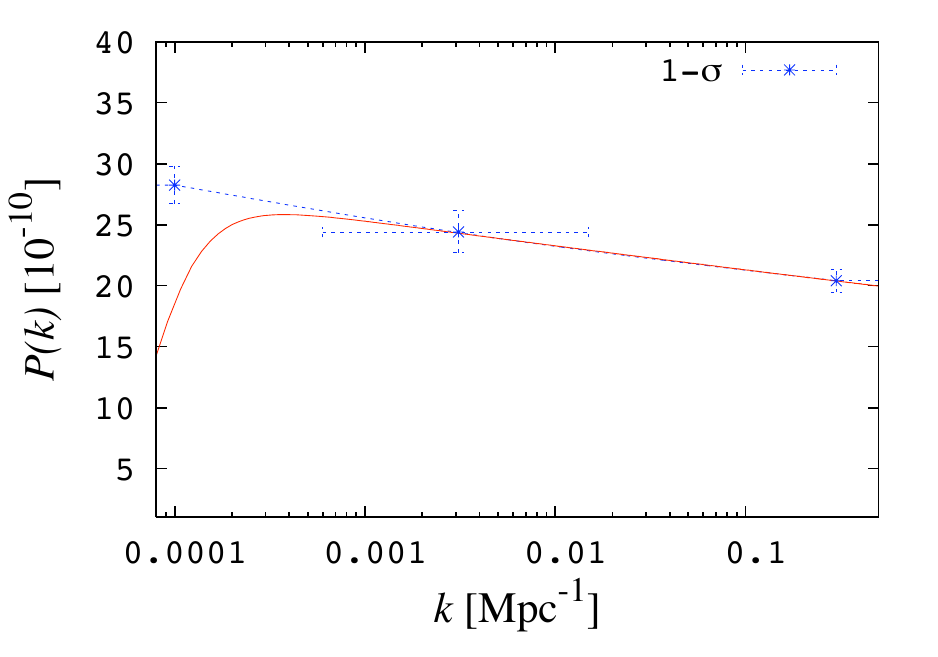}
\end{array}$
\end{center}
\caption{Reconstruction of the Lasenby  \&  Doran primordial spectrum  based  on the binning  format
with cubic  spline described
in Section \ref{sec;k_i}. We have  assumed  an idealised  CMB spectrum with limitation only  due  cosmic variance (left).
Right panel shows  the  reconstructed spectrum in the  binning  format  together  with  the LD input spectrum.}
\label{fig:LD_noise2}
\end{figure}

In this appendix we illustrate how the type of interpolation in our
node-based approach can influence the reconstruction of the primordial
spectrum.  We use the same example shown in Figure \ref{fig:LD_noise}
but now we use a cubic spline to interpolate through the $k$-nodes.
From Figure \ref{fig:LD_noise2} we note that 
the spectrum used as an input lies well outside the error bar on the reconstruction at  low
$k$-values.
Therefore, the spline fails  to recover the 
 input spectrum correctly, contrary to the linear interpolation where the
 recovered  spectrum, shown in Figure \ref{fig:LD_noise}, is  more representative,
 with the true spectrum lying comfortably within the error-bars on the reconstruction at all $k$-values.
 This is mainly
because a function with rapidly changing higher derivatives, such as the
input primordial spectrum used here, is less accurately approximated
by higher order polynomials. 
In particular, the requirement of continuous first and second derivatives, combined with 
the tight constraints at small and intermediate length scales, leads  to a significant
overestimation of the power at the less  well  constrained region  at the largest  scales.
 Hence, in this case, the linear
interpolation describes the shape of the primordial spectrum better
than the spline. 
\\

\begin{figure}
\begin{center}$
\begin{array}{cc}
  \mathcal{B}_{3,1} =+2.90 \pm 0.30  & \mathcal{B}_{4,1} =+1.62 \pm 0.30  \\
\includegraphics[trim = 1mm  -2mm 5mm -5mm, clip, width=4.5cm, height=4.cm]{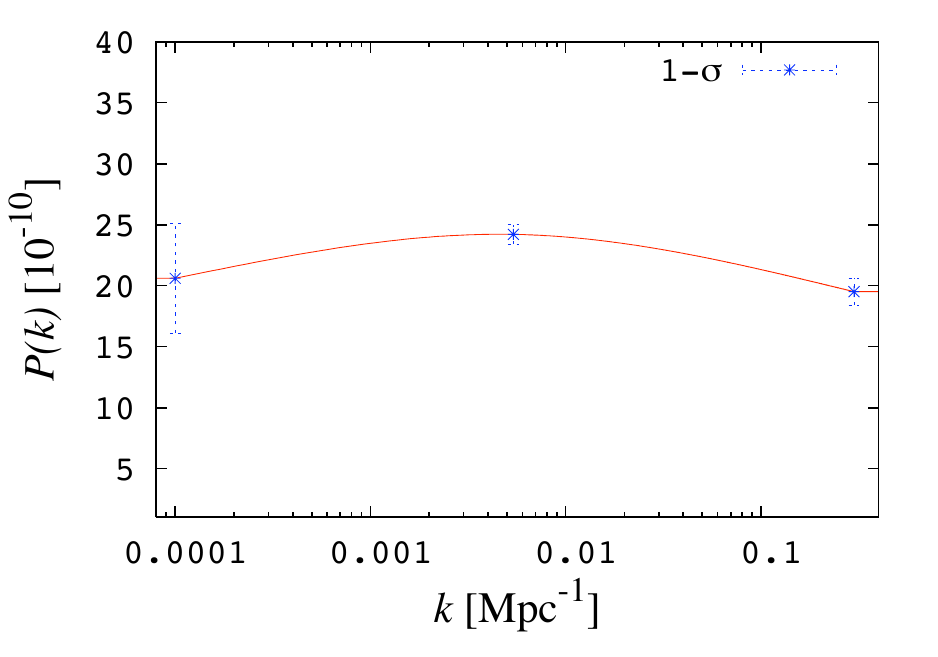} 
\includegraphics[trim = 45mm  95mm 50mm 95mm, clip, width=3.5cm, height=4.cm]{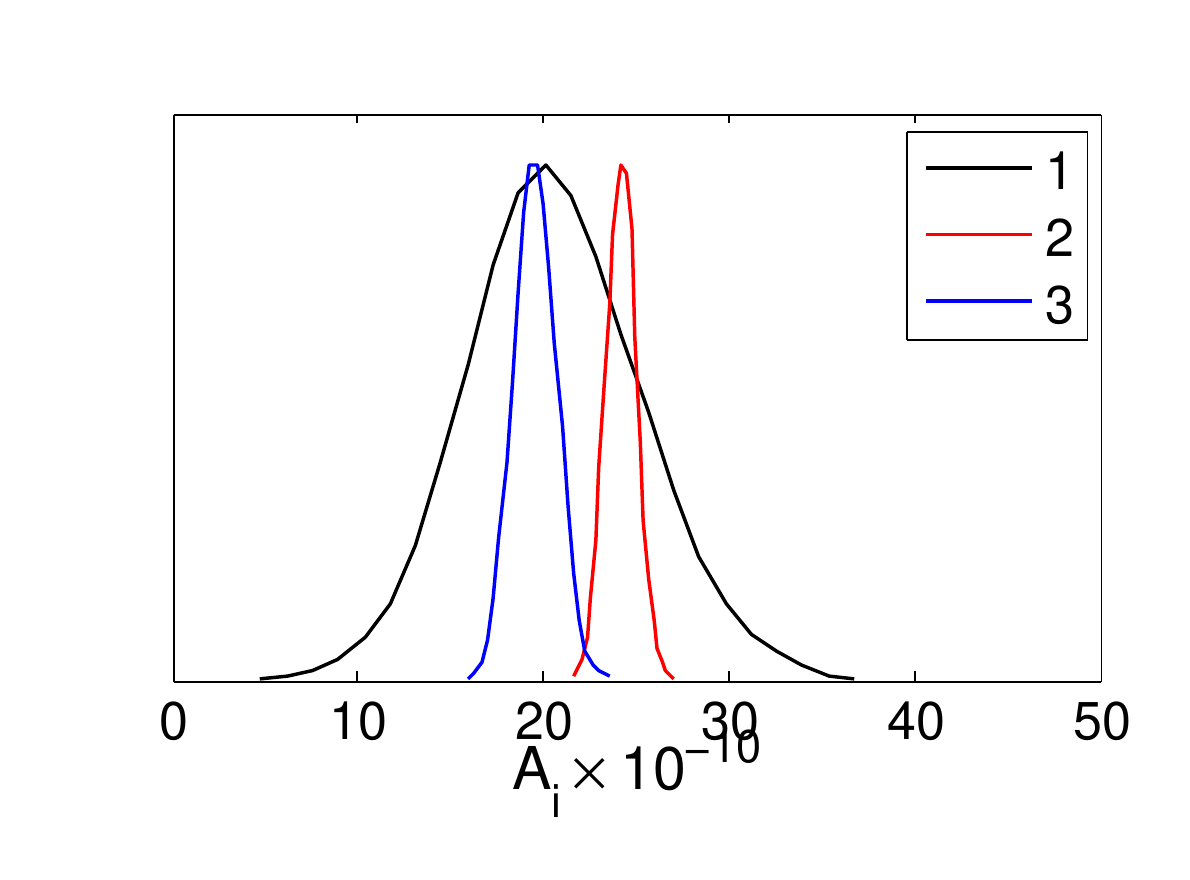}&
\includegraphics[trim = 1mm  -2mm 5mm -5mm, clip, width=4.5cm, height=4.cm]{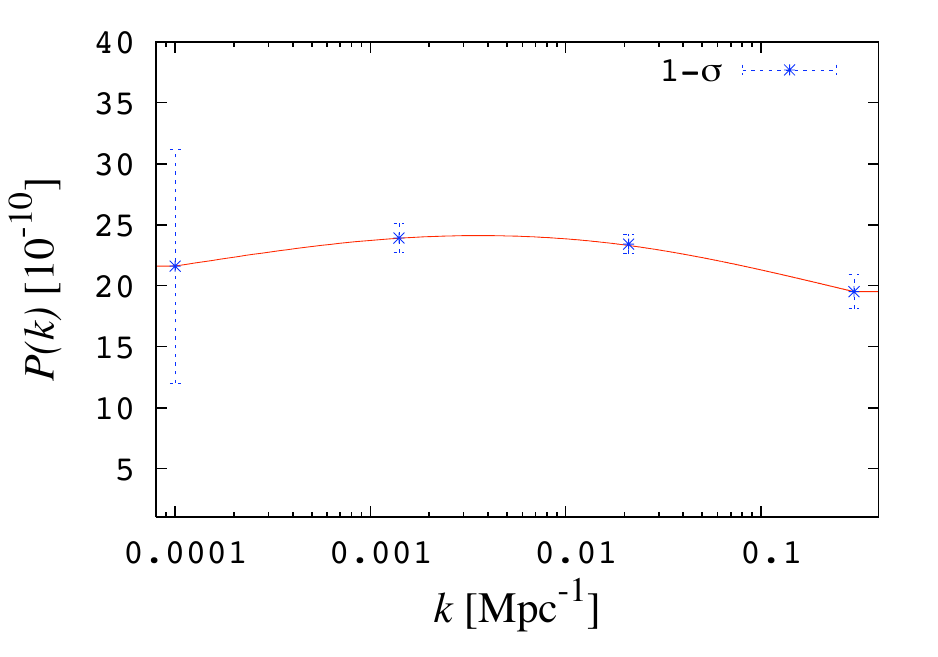} 
\includegraphics[trim = 45mm  95mm 50mm 95mm, clip, width=3.5cm, height=4.cm]{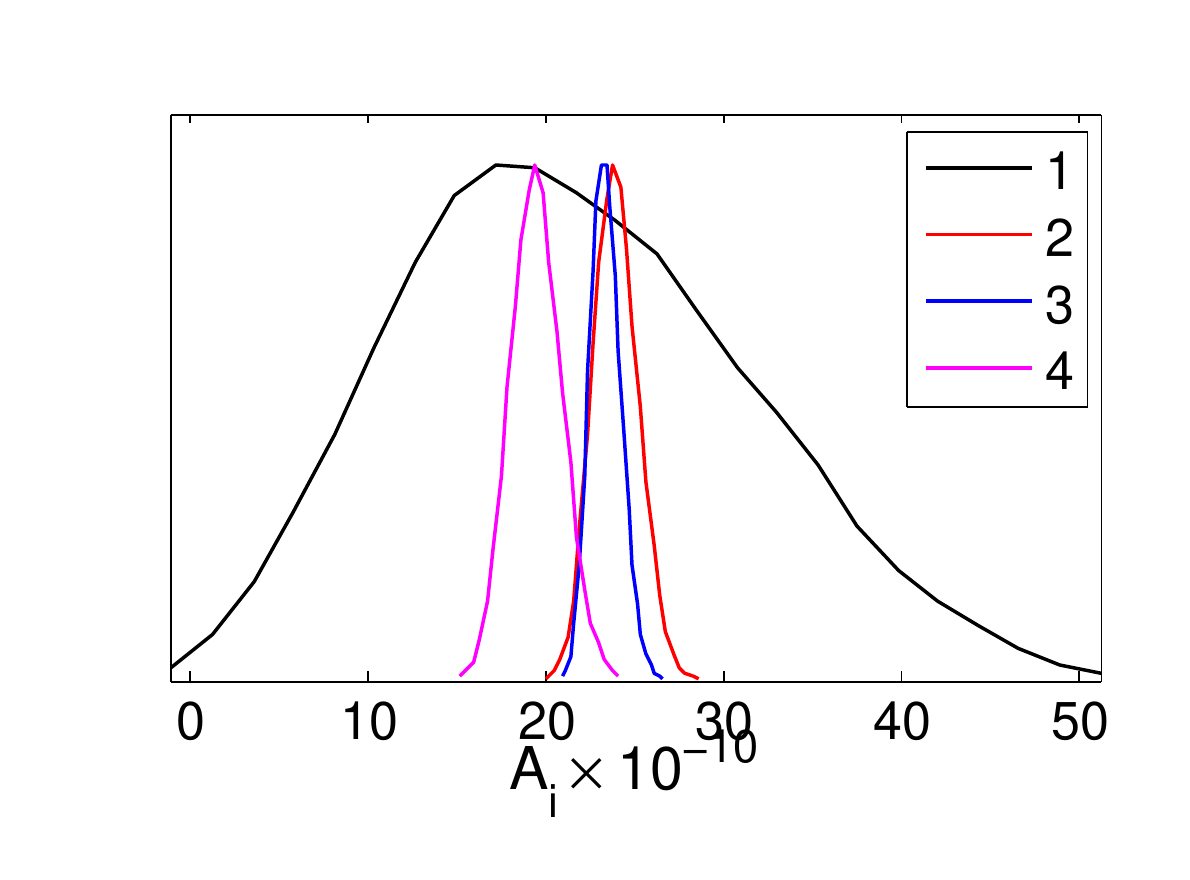} \\
 $$\mathcal{B}_{k_1,1} =+4.21 \pm 0.30$$ &  $$\mathcal{B}_{k_2,1} =+4.02 \pm 0.30$$ \\
 \includegraphics[trim = 1mm  -2mm 5mm -5mm, clip, width=4.5cm, height=4cm]{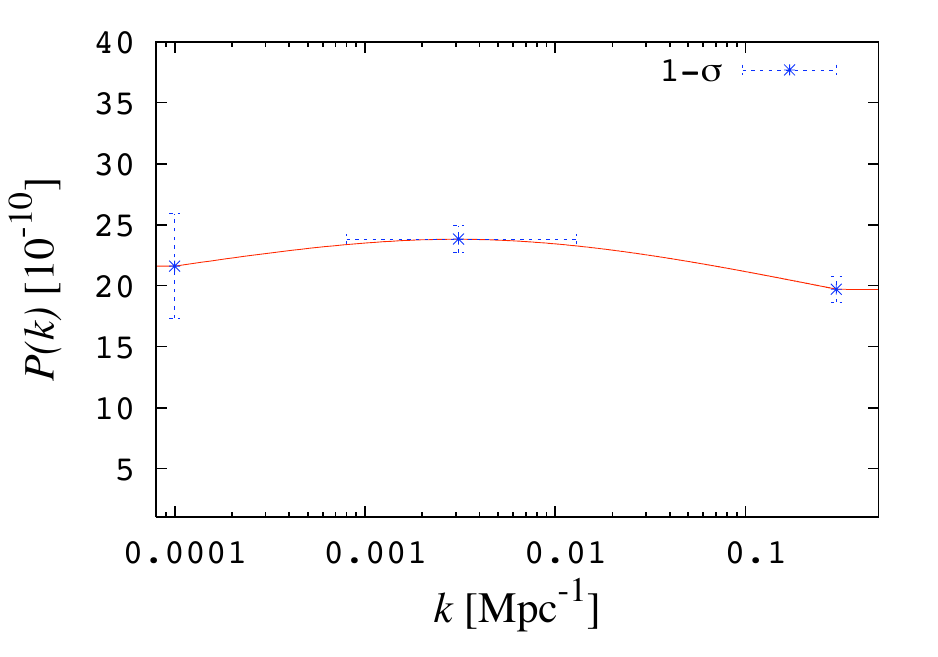}
\includegraphics[trim = 32mm  52mm 40mm 62mm, clip, width=3.5cm, height=4cm]{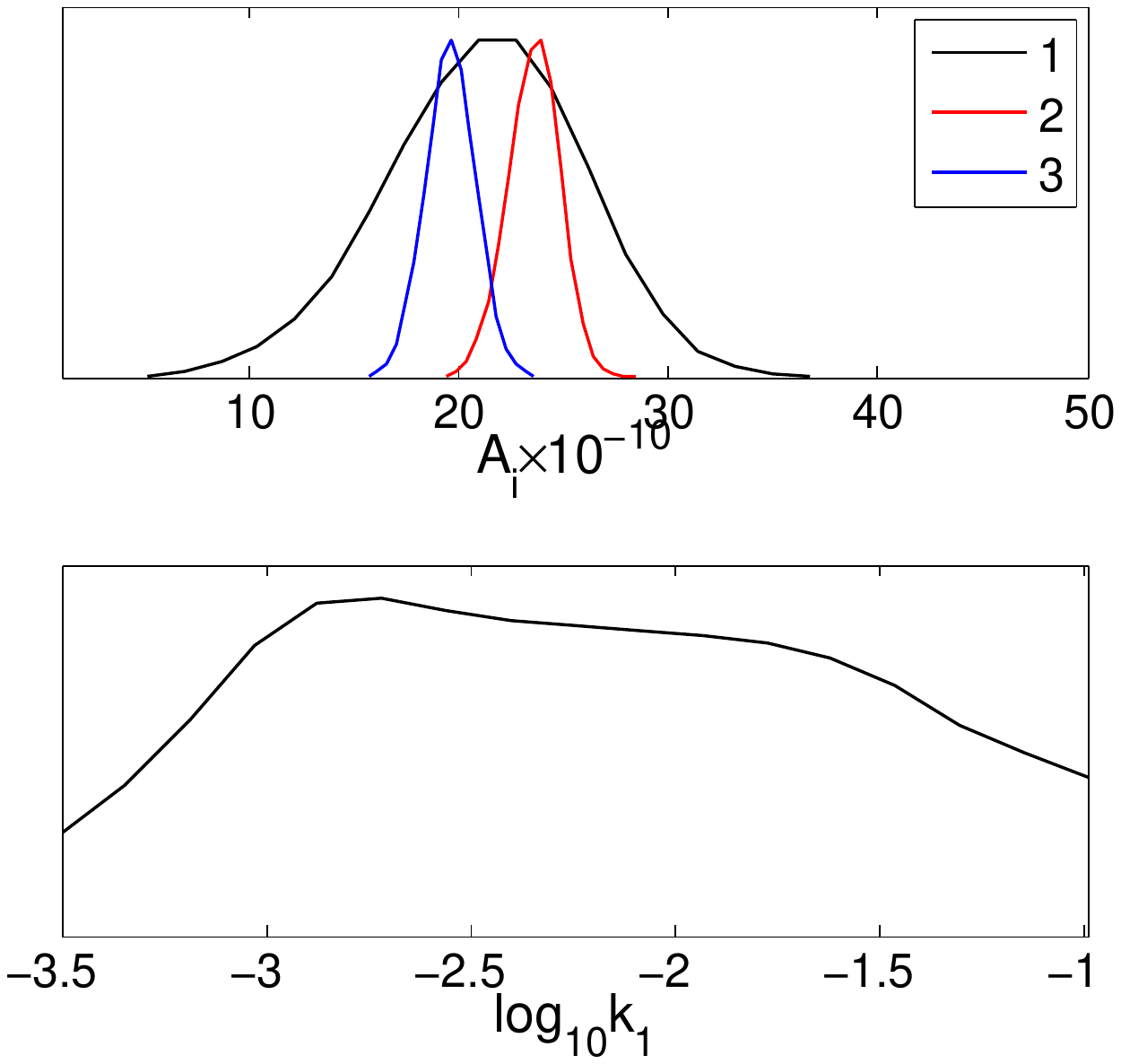} &
 \includegraphics[trim = 1mm  -2mm 5mm -5mm, clip, width=4.5cm, height=4cm]{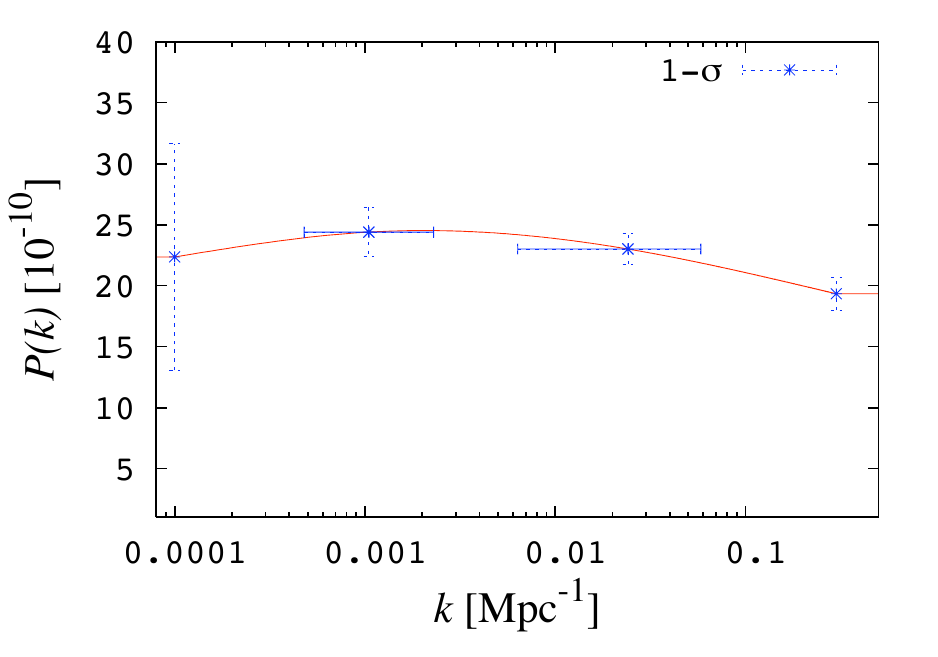}
\includegraphics[trim = 32mm  52mm 40mm 62mm, clip, width=3.5cm, height=4cm]{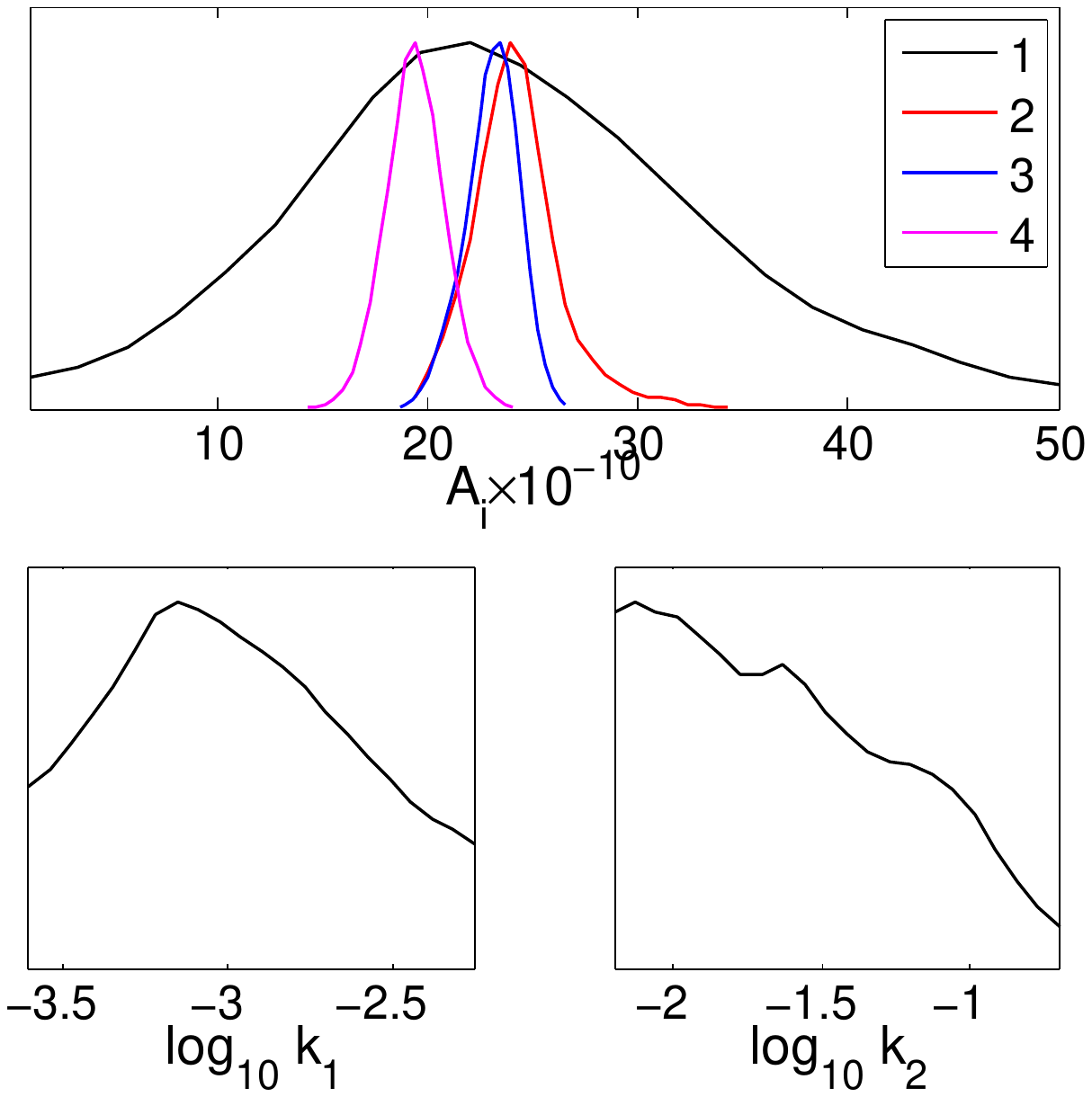} \\
 $$\mathcal{B}_{k_3,1} =+3.81 \pm 0.30 $$  \\
\includegraphics[trim = 1mm  -2mm 5mm -5mm, clip, width=4.5cm, height=4cm]{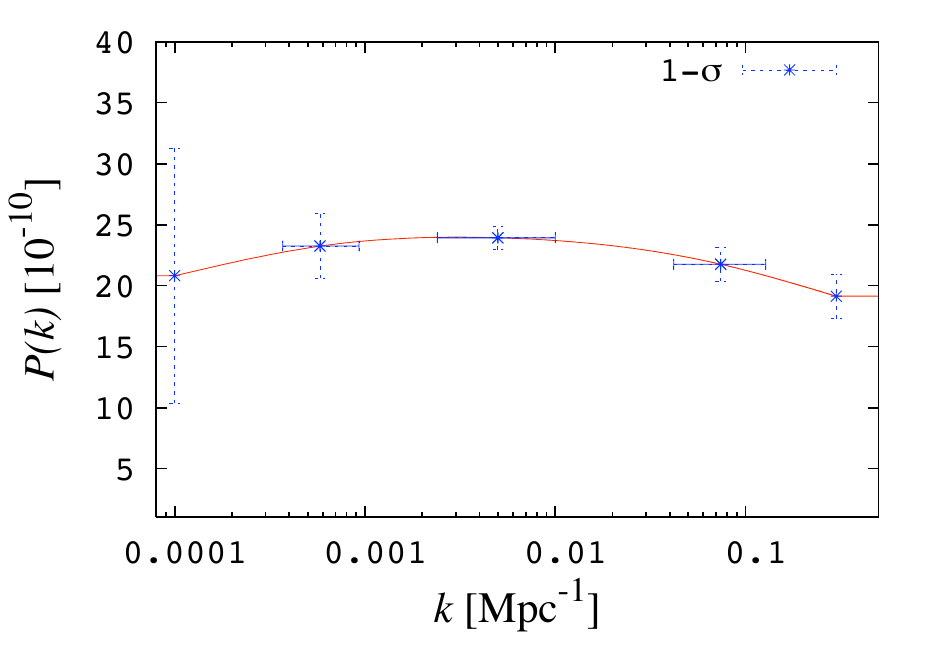}
\includegraphics[trim = 32mm  75mm 30mm 82mm, clip, width=4.0cm, height=4cm]{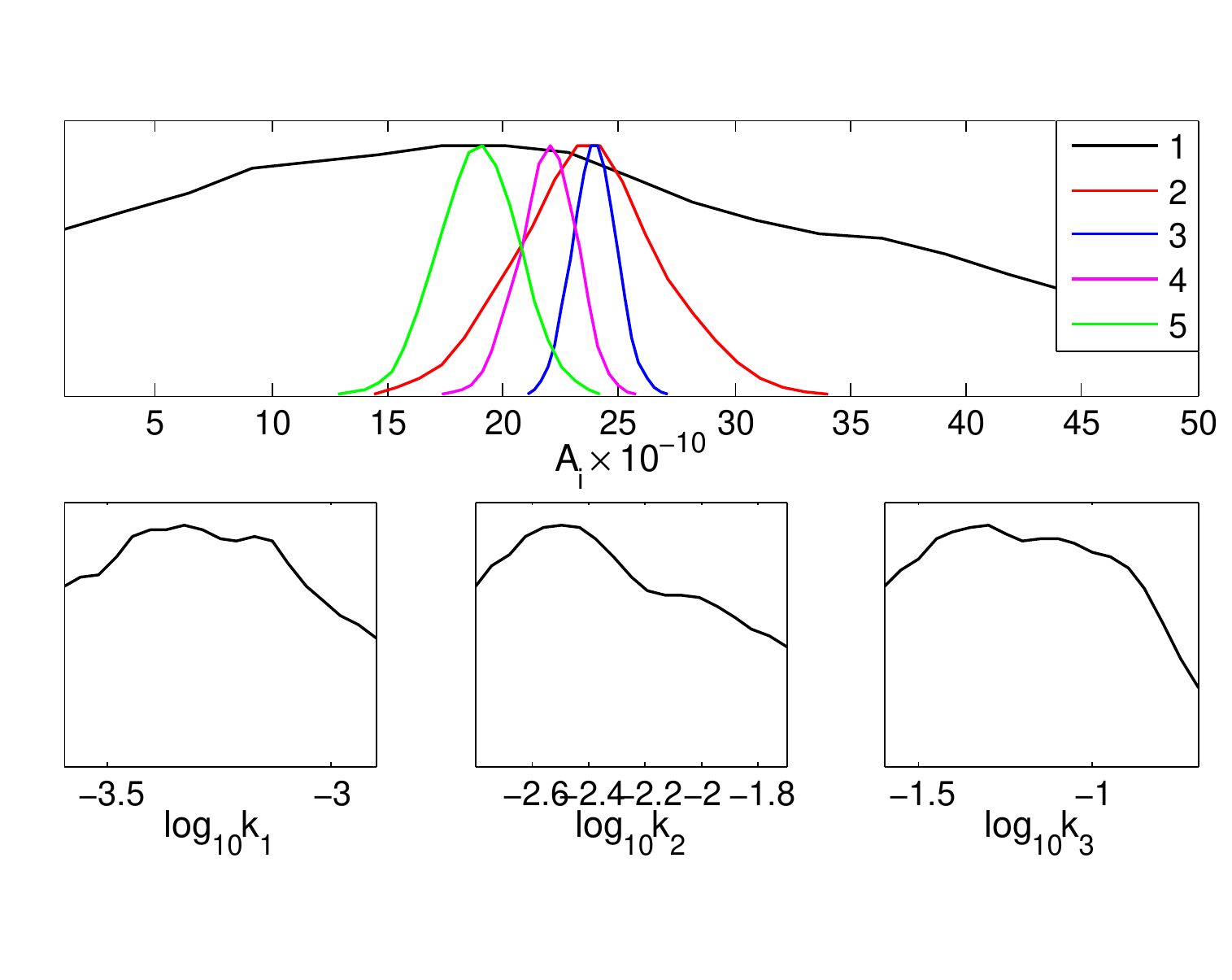} 
\end{array}$
\end{center}
\caption{Reconstruction of the primordial spectrum using the cubic spline.
 Top panel resembles plots  (b) and (c) shown in  Figure \ref{fig:recons_1}, whereas
 bottom panel the  reconstruction for the models shown  in  Figure \ref{fig:mov_k}.   
 To the  right  of each reconstruction we plot the
 1D marginalised posterior distribution of the amplitudes $A_i$ and $k$-node position $k_i$.
   The top label  in each  panel  denotes the associated Bayes factor with  respect
to the base model (HZ) shown in Figure \ref{fig:recons_1} (a).}
\label{fig:recons_2}
\end{figure}

For comparison with the results presented earlier, we also use the
cubic spline to perform similar node-based reconstructions of the
spectrum to those shown in Figures \ref{fig:recons_1} and \ref{fig:mov_k}. The
low number of bins used to describe the global structure of the
spectrum yield to similar shapes by using both interpolation methods.  
The reconstructed spectra for three and four bins, along  with one and two 
internal $k$-nodes, are plotted in Figure \ref{fig:recons_2} using the spline interpolation.

\end{document}